\documentclass[journal,twoside,10pt]{IEEEtran}
\pdfoutput=1
\usepackage{latexsym, graphicx}
\usepackage{amsmath,cases}
\usepackage{color}
\usepackage{algorithm,algpseudocode,float}
\usepackage{cite}
\usepackage{amsmath,amssymb,amsfonts}
\usepackage{latexsym, graphicx}
\usepackage{amsmath,cases}
\usepackage{algorithm,algpseudocode,float}
\usepackage{tabularx}
\usepackage{booktabs}
\usepackage{graphicx}
\usepackage{textcomp} 
\usepackage{multirow}
\usepackage{epstopdf} 

\usepackage{cuted}
\usepackage{bm}
\usepackage{mathtools}
\usepackage{lipsum,mwe,cuted}
\usepackage{float}
\usepackage{stfloats}

\makeatletter
\newenvironment{breakablealgorithm}
{
		\begin{center}
			\refstepcounter{algorithm}
			\hrule height.8pt depth0pt \kern2pt
			\renewcommand{\caption}[2][\relax]{
				{\raggedright\textbf{\ALG@name~\thealgorithm} ##2\par}%
				\ifx\relax##1\relax 
				\addcontentsline{loa}{algorithm}{\protect\numberline{\thealgorithm}##2}%
				\else 
				\addcontentsline{loa}{algorithm}{\protect\numberline{\thealgorithm}##1}%
				\fi
				\kern2pt\hrule\kern2pt
			}
		}{
		\kern2pt\hrule\relax
	\end{center}
}
\makeatother

\begin{document}
	
	
	\title{\textcolor{black}{Federated Multi-Agent Deep Reinforcement Learning Approach via Physics-Informed Reward for Multi-Microgrid Energy Management}}
	
	\author{Yuanzheng Li, {\it Member IEEE}, Shangyang He, Yang Li, {\it Senior Member IEEE}, \\ Yang Shi, {\it Fellow IEEE}, 
		and Zhigang Zeng, {\it Fellow IEEE} 
		
		
		
		\thanks{This work is supported in part by the National Natural Science Foundation of China (Grant 62073148), in part by Key Project of National Natural Science Foundation of China (Grant 62233006), and in part by Key Scientific and Technological Research Project of State Grid Corporation of China (Grant No. 1400-202099523A-0-0-00). (Corresponding author: Yang Li)}
		
		\thanks{Y. Z. Li and Z. G. Zeng are with School of Artificial Intelligence and Automation, Key Laboratory on Image Information Processing and Intelligent Control of Ministry of Education, Huazhong University of Science and Technology, Wuhan 430074, and also with China-Belt and Road Joint Laboratory on Measurement and Control Technology, Wuhan, China, 430074 (Email: Yuanzheng\_Li@hust.edu.cn, zgzeng@hust.edu.cn).}
		
		\thanks{S. Y. He is with China-EU Institute for Clean and Renewable Energy, Huazhong University of Science and Technology, Wuhan 430074, China (Email: heshangyang10@hust.edu.cn).}
		
		\thanks{Y. Li is with School of Electrical Engineering, Northeast Electric Power University, Jilin 132012, China (Email:liyang@neepu.edu.cn).}
		
		
		\thanks{Y. Shi is with the Department of Mechanical Engineering, University of Victoria, Victoria, BC V8P 5C2, Canada (E-mail:yshi@uvic.ca).}
		
	}
	
	\maketitle
	
	\begin{abstract}
		The utilization of large-scale distributed renewable energy promotes the development of the multi-microgrid (MMG), which raises the need of developing an effective energy management method to \textcolor{black}{minimize economic costs and keep self energy-sufficiency}. The multi-agent deep reinforcement learning (MADRL) has been widely used for the energy management problem because of its real-time scheduling ability. \textcolor{black}{However, its training requires massive energy operation data of microgrids (MGs), while gathering these data from different MGs would threaten their privacy and data security.} \textcolor{black}{Therefore, this paper tackles this practical yet challenging issue by proposing a federated multi-agent deep reinforcement learning (F-MADRL) algorithm via the physics-informed reward.} In this algorithm, the federated learning (FL) mechanism is introduced to train the F-MADRL algorithm thus ensures the privacy and the security of data. \textcolor{black}{In addition, a decentralized MMG model is built, and the energy of each participated MG is managed by an agent, which aims to minimize economic costs and keep self energy-sufficiency according to the physics-informed reward.} 
		At first, MGs individually execute the self-training based on local energy operation data to train their local agent models. Then, these local models are periodically uploaded to a server and their parameters are aggregated to build a global agent, which will be broadcasted to MGs and replace their local agents. In this way, the experience of each MG agent can be shared and the energy operation data is not explicitly transmitted, thus protecting the privacy and ensuring data security. \textcolor{black}{Finally, experiments are conducted on Oak Ridge national laboratory distributed energy control communication lab microgrid (ORNL-MG) test system, and the comparisons are carried out to verify the effectiveness of introducing the FL mechanism and the outperformance of our proposed F-MADRL.}
	\end{abstract}
	\begin{IEEEkeywords}
		Multi-microgrid, multi-agent deep reinforcement learning, federated learning, proximal policy optimization.
	\end{IEEEkeywords}
	
	\section{Introduction}
	In recent years, renewable energy (RE) has been widely deployed, such as wind power and photovoltaic. Unlike traditional power plants, RE resources are usually distributed. Therefore, microgrids (MGs) have been paid much attention to utilize the RE. Note that the MG usually works in a local area, and provides the required electricity for a small entity, such as a school, a hospital, or a community \cite{SunQY1,SunQY2,ZhangHF,Add1}.
	
	Normally, the main target of MG is to achieve the self-sufficiency of energy via the utilization of RE. However, due to its limited capacity, the MG has to take the risk of power shortage. Specifically, since the user demand and RE are dependent on the user behavior and weather condition, the power demand may exceed the capacity of MG while RE generation may be insufficient, thus causing the power shortage \cite{P0-Distributed3}. For this reason, numerous adjacent MGs are interconnected to form a multi-microgrid (MMG) system. \textcolor{black}{Compared with an isolated MG, the MMG system is more capable of utilizing RE because of its larger capacity. Besides, although these MGs belong to different entities, the energy is allowed to be traded among different MGs, i.e., each MG can actively sell its surplus power when its power generation exceeds the demand, or purchase power from other MGs when the generation is insufficient \cite{P1-isolateMG}.}
	Therefore, the MMG is more promising to achieve energy self-sufficiency compared with an isolated MG. 
	
	However, because of the complexity of energy management of the MMG, it is essential to adopt an effective scheme. The present studies of MMG energy management can be mainly categorized into two types, i.e., the centralized and decentralized schemes. The former one is based on a centralized energy management center, which could get access to the related energy information of all MGs in the MMG system \cite{CMMG01, Add3}. Then, this center can well make decisions to achieve the energy self-sufficiency of the MMG system. However, note that the multiple MGs usually belong to different entities, and it is difficult for the centralized management center to acquire operation data of all MGs due to the increasing awareness of privacy protection. 
	
	Therefore, a more popular research direction is the decentralized MMG management scheme. For instance, Ng \emph{et al}. have proposed the concept of MMG control, which uses the multi-agent approach to achieve the decentralized control of each MG \cite{FirstMMGControl}. In addition, Yang \emph{et al}. have adopted multiple self-decision agents replacing the energy management center for the energy self-sufficiency of participated MGs \cite{MAMMG_01}. Liu \emph{et al}. have treated the MMG system as a fully distributed optimization model, which is solved by a robust optimal scheduling algorithm \cite{LiuYunDMMG}. Moreover, Ref. \cite{MAMMG_03} has proposed a multi-agent MMG system, where the individual agent of each MG collects the data from local units and performs optimization separately. In addition, Ref. \cite{MAMMG_04} has proposed the MG agents to well utilize the partially observed information, for achieving the optimal energy management of MMG. 
	
	\textcolor{black}{The aforementioned literature focuses on building accurate optimization models, which can be summarized as the model-based approach. However, there exists an essential drawback, i.e., the model-based approach is merely suited for the predetermined scheduling solution rather than a real-time decision. In other words, the predetermined scheduling is difficult to handle emergencies or the unexpected change in the load demand occurring in the MMG system.}
	
	To tackle this problem, the learning-based approach has been developed in recent years \cite{Add2, Add4}. \textcolor{black}{Benefiting from the development of the physics-informed deep learning techniques, the outputs of the black-box model are more generalized and interpretable \cite{psysic_inform}.}
	One of the most representative approaches is multi-agent deep reinforcement learning (MADRL), which is widely deployed in the MMG energy management problem due to its nature of interacting with the physical characteristic of the real world \cite{RL_MMG_0, Friend}.
	For instance, The MADRL used in Ref. \cite{RL_MMG_6} observes the temperature, energy generation and other physical parameters to control soft load and transaction effectively for MMG. The experiments demonstrate the convergence of these algorithms and emphasize the outperformance of the actor-critic algorithm. 
	Ref. \cite{RL_MMG_8} proposes an energy management approach that takes advantage of a multi-agent model-free reinforcement learning algorithm. This distributed and hierarchical decision mechanism effectively increases the energy self-sufficiency of MMG. 
	\textcolor{black}{Besides, a MADRL method is adopted in Ref. \cite{RL_MMG_3} to realize the post-disaster resilience of distributed MG system.} Aiming to increase the income of the system, the MADRL shows its strong adaptability in different conditions through experiments.
	Moreover, the implementation of MADRL would significantly increase the autonomy of each MG \cite{RL_MMG_7}\cite{RL_MMG_4}. \textcolor{black}{For instance, a MADRL framework based on the deep neural network is proposed in Ref. \cite{RL_MMG_7} to improve the operational performance and autonomy of each participant MG.} Ref. \cite{RL_MMG_4} sets the agents in different MGs for the distributed control and achieve higher MG autonomy. To balance the benefits of the MMG participants and guarantee the efficiency, Ref. \cite{RL_MMG_5} proposes an equilibrium selection multi-agent reinforcement learning algorithm based on Q-learning to promote the autonomy of MG operation.
	
	However, since the MADRL technology requires massive data to train the MG agent, the concern of user privacy is raised. The data of the users can be utilized to analyze their habits and even their life tracks. In the case of MMG energy management, to train an effective agent with a high generalization, massive energy operation data should be collected from different MGs. \textcolor{black}{However, although each MG aims to pursure a better performance through experiences sharing, they may be not willing to submit their processing data because of privacy awareness \cite{RL_MMG_7}. On the other hand, the security during data transmission cannot be guaranteed.}
	
	
	\textcolor{black}{To tackle the above issues, we introduce an emerging distributed learning approach, namely federated learning (FL), for training MADRL in the MMG energy management via physics-informed reward \cite{FL_IdeaSource}.} In other words, we apply the FL to protect user privacy and guarantee data security while ensuring the generalization of each MG agent in the MMG system. Specifically, each MG is controlled by an agent, which deploys a recent deep reinforcement model, namely proximal policy optimization (PPO) \cite{PPO}. Each agent firstly executes the self-training according to the local energy operation data of each MG to maximize the physics-informed reward, i.e., the economic operation and self energy-sufficiency.
	Then, the agents upload their local model parameters, such as the weights and biases of the model, to a server. After that, these parameters are aggregated by the server to construct a global model, which will be broadcasted to each MG and replace the local model. In this way, agents share their experiences through the FL mechanism, which thus enhances their generalization\footnote{\textcolor{black}{Since the MGs in the MMG belong to different kinds of entities, their local operation data manifest the perference of local users. Thus the agent trained by local data would be confronted of performances decline when operating in other MGs, and this phenoma is termed as the generalization decrease of the MG agent.}} compared with the local training. Moreover, the FL mechanism only requires model parameters, and the operation data of each MG would stay locally. Therefore, the user privacy and data security can be guaranteed.
	
	
	The main contributions of this paper are presented as follows.
	
	\textcolor{black}{(1) A MMG system model is developed for the deployment of FL, where each MG contains conventional generators (CGs), batteries (BAs), renewable energy generators (REGs), load and the energy management center. Then, a server is introduced to implement the FL mechanism which can communicate with MGs and aggregates the parameters of the MG agent, such as the weight and bias of the neural network models. Since the server would not perform as a center of MMG that guides the decisions of each MG, MGs would endow a high autonomy and suffer from less risk of privacy leakage.}
	
	(2) \textcolor{black}{A federated multi-agent deep reinforcement learning algorithm (F-MADRL) is proposed for the energy management of the MMG system. Each MG has an agent that collects the operation data for self-training. Then, the agent parameters are uploaded to the server and aggregated to a global agent. Afterwards, the agent of each MG is replaced by the global one. In this way, the privacy of each MG user can be protected.}
	
	(3) A physics-informed reward is developed by orienting targets of the MG agent, i.e., the economic operation and the self energy-sufficiency. \textcolor{black}{The MG agents trained through the physics-informed reward would be endowed with a better interpretation of action because of the consideration of physical targets.}
	
	(4) \textcolor{black}{Case studies conducted on the Oak Ridge national laboratory distributed energy control communication lab microgrid (ORNL-MG) test system \cite{ORNL_MG_Fig} demonstrate that our proposed F-MADRL algorithm is effective under different demands and renewable energy scenarios.} Moreover, we verify that F-MADRL outperforms other state-of-the-art DRL algorithms under the distributed MMG model. 
	
	The remainder of this paper is organized as follows. Section II introduces the theoretical basis of the reinforcement learning. In Section III, a decentralized MMG model is built. Section IV proposes the F-MADRL algorithm, and provides its overall structure and technical details. In Section V, comprehensive case studies are conducted. Finally, Section VI concludes this paper.
	
	\section{Theoretical Basis of Reinforcement Learning}
	\textcolor{black}{Normally, the Markov decision process (MDP) is defined by a five-tuple $\langle\mathcal{S},\mathcal{A},\mathcal{P},\mathcal{R},\gamma\rangle$, where $\mathcal{S}$ is the finite state space that stands for all valid states and $\mathcal{A}$ represents the finite set of actions. $\mathcal{P}=\{p(s_{t+1}|s_t, a_t)\}$ stands for the set of transition probability from state $s_t$ to $s_{t+1}$, and $\mathcal{R}=r(s_t,a_t), \mathcal{R}\in\mathbb{R}$; $\mathcal{S\times A}\rightarrow \mathbb{R}$ is termed as the reward function, which is normally the metric to evaluate the action. $\gamma\in[0,1]$ indicates the discount factor, which represents the importance of the present reward \cite{Friend, Guokai}.}
	
	To solve the MDP, a policy $\pi$ should be developed to provide the probability of executing action $a$ when observing the state $s$, i.e. $\pi (a|s) = P[A_t =a|S_t=s]$. The aim of $\pi$ is to maximize the discounted cumulative reward during the finite time $T$, which is termed as the return function:
	{\color{black}
		\begin{equation}
			\small
			\setlength{\abovedisplayskip}{3pt}
			\setlength{\belowdisplayskip}{3pt}
			U_t =  \sum_{k=t}^{T}\gamma^{k-t} r(s_{k},a_{k})
		\end{equation}
		where $r(s_{k},a_{k})$ is the reward function, which calculates the reward value under state $s_k$ with action $a_{k}$; $\gamma\in[0,1]$ is the discount factor, representing the importance of the future reward \cite{sutton2018reinforcement}. Then, two kinds of value functions are defined based on $U_t$ to help the policy make decisions.} The first is the state value function $V_{\pi}(s)$ and the other is the action value function $Q_{\pi}(a,s)$, which are formulated as follows:
	\begin{equation}
		\small
		\begin{split}
			V_{\pi}(s)&=E_{\pi}[U_t|S_t =s]\\
			&=\sum_{a}\pi (a|s)\sum_{s'}P^a_{ss'}[r(s,a)+\gamma V_{\pi}(s^{'})]\label{eq:state_value}
		\end{split}	
	\end{equation}
	\begin{equation}
		\small
		\color{black}
		\begin{split}
			Q_{\pi}(a,s)&=E_{\pi}[U_t|S_t =s, A_t=a]\\
			&=\sum_{s'}P^a_{ss'}[r(s,s'|a)+\gamma\sum_{a'}Q_{\pi}(a',s')]
		\end{split}
	\end{equation}
	where $V_{\pi}(s)$ stands for the expectation of future reward at the state $s$, and the $Q_{\pi}(a,s)$ represents the future expected reward when selecting an action $a$ at state $s$. $s'$ and $a'$ stand for the possible reaching state and action at state $s$. $P^a_{ss'}$ is the transition probability from $s$ to $s'$ under $a$. In fact, $V_{\pi}(s)$ and $Q_{\pi}(a,s)$ are used to evaluate the quality of the state $s$ and the action-state pair $(a,s)$, respectively. They are updated according to above two equations and help the policy $\pi$ decide whether reaching the state or executing the action. 
	
	\section{The Decentralized Multi-Microgrid Energy Management Model}
	The decentralized MMG system includes numerous MGs that are connected to a distribution power network. Usually, an energy management center is set in each MG, which performs as an agent to conduct self-training and control the dispatchable elements, such as conventional generators (CGs), batteries (BAs), etc. In this section, to describe the energy management model of the MMG system more clearly, we firstly introduce the isolated MG model with the MDP format before developing the MMG model.
	
	\subsection{The Isolated Microgrid Energy Management Model}
\begin{figure}[bhtp]
		\centering
		\includegraphics[width=3in]{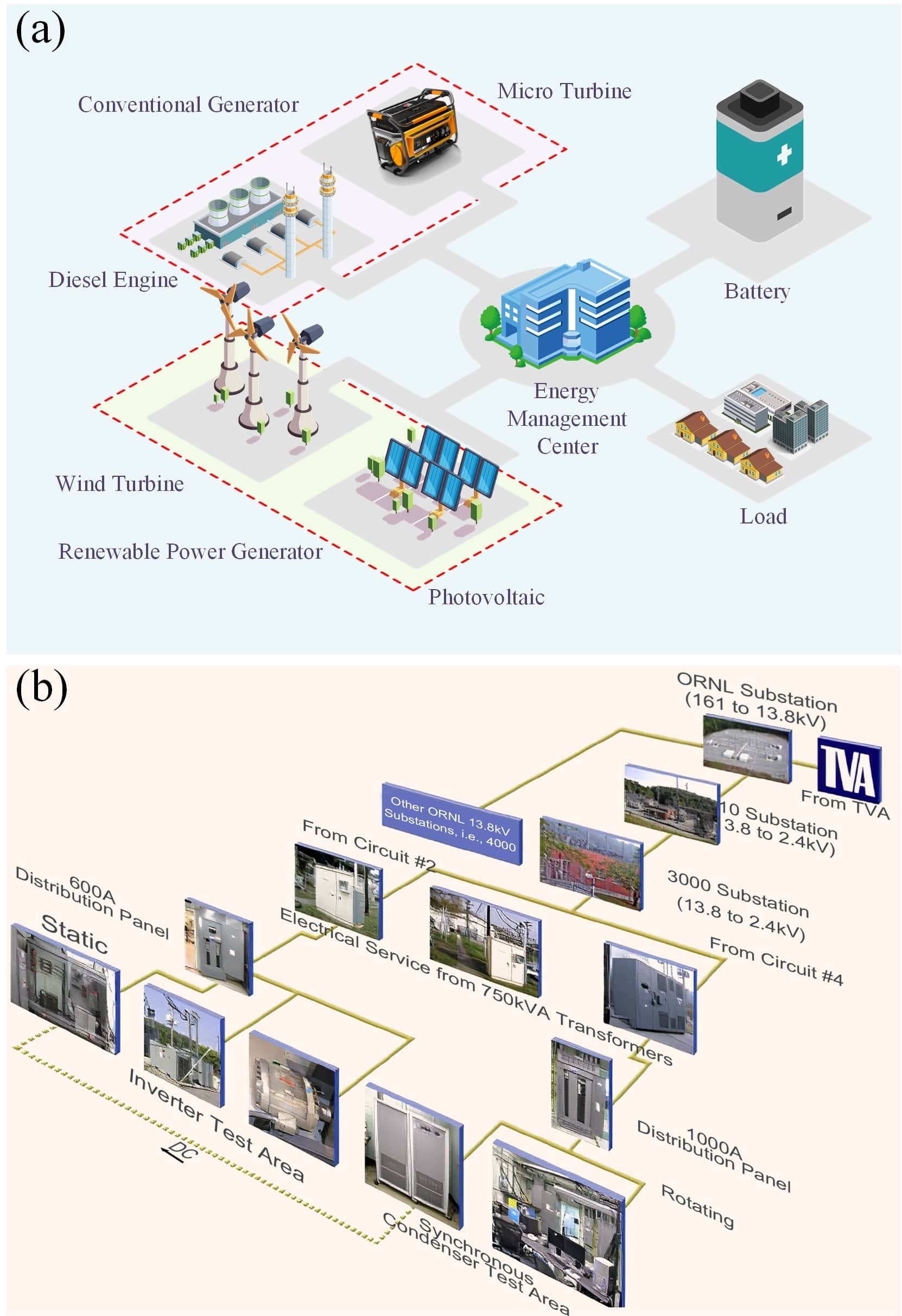}
		\caption{\textcolor{black}{The structure of (a) an isolated MG and (b) the ORNL-MG \cite{ORNL_MG_Fig}.}}\label{fig:singleMG}
		\vspace{-0.2cm}
	\end{figure}

	\textcolor{black}{Fig. \ref{fig:singleMG} illustrates the structure of the isolated MG model and a real-world MG system case. Normally, as shown in Fig. \ref{fig:singleMG}(a), a MG is constructed by five types of elements: renewable power generators, BA, CG, conventional load (CL) and energy management center. Note that BAs and CGs are dispatchable since their outputs are controlled by the management center. On the contrary, because of the high uncertainties of RE, \textcolor{black}{the outputs of REG cannot be controlled.} Additionally, the energy management center is termed as the agent that controls these dispatchable elements by observing the state of MG operation. Following this structure, the Oak Ridge national laboratory distributed energy control communication lab microgrid test system (ORNL-MG) is selected as the real-world case in this paper, which is illustrated in Fig. \ref{fig:singleMG}(b).}
	\subsubsection{Conventional Generator}
	It can be seen from Fig. \ref{fig:singleMG} that the CG includes diesel engine generator and micro turbine, which generate power through fossil fuels. The cost functions of CGs can be represented as follows:
	\begin{equation}
		\small
		\setlength{\abovedisplayskip}{3pt}
		\setlength{\belowdisplayskip}{3pt}
		C(P_{\text{CG},i})=a_{\text{CG},i} P_{\text{CG},i}^2 + b_{\text{CG},i} P_{\text{CG},i} + c_{\text{CG},i}
	\end{equation}
	\begin{equation}
		\small
		\setlength{\abovedisplayskip}{3pt}
		\setlength{\belowdisplayskip}{3pt}
		P_{\text{CG},i}^{\text{min}}\leq P_{\text{CG},i} \leq P_{\text{CG},i}^{\text{max}}\label{eq:constrain_CG}
	\end{equation}
	where $C(P_{\text{CG},i})$ represents the generation cost of $i$th CG, and $P_{\text{CG},i}$ is its generation power. \textcolor{black}{$a_{\text{CG},i}$, $b_{\text{CG},i}$ and $c_{\text{CG},i}$ denote the cost coefficients of the $i$th CG. $P_{\text{CG},i}^{\text{min}}$ and $P_{\text{CG},i}^{\text{max}}$ are the lower and upper bounds of the $i$th CG.}
	
	\subsubsection{\textcolor{black}{Renewable Energy Generator}}
	\textcolor{black}{Fig. \ref{fig:singleMG} presents two kinds of REGs, namely wind turbine and photovoltaic. The generation of REG normally depends on the natural environment such as wind speed, temperature, weather and solar irradiance \cite{ZhuoLing}.}
	\textcolor{black}{Since the REGs do not consume any fossil fuels, their generation costs are not considered in this paper.}
	
	\subsubsection{Battery}
	As one of the most commonly used energy storage devices, \textcolor{black}{BA can store energy generated by CGs and REGs, and release it when needed.} Thus, the BA has two operation states, namely charging and discharging, which are represented by the transition of its state of charge, and can be formulated as follows:
	{\color{black}
		\begin{eqnarray}
			\small
			\setlength{\abovedisplayskip}{3pt}
			\setlength{\belowdisplayskip}{3pt}
			SOC^{t+1} = (1-\delta)SOC^t-\frac{P_{\text{BA}}^t}{\eta_{\text{ch}}C_{\text{BA}}}\\
			SOC^{t+1} = (1-\delta)SOC^t-\frac{\eta_{\text{dch}}P_{\text{BA}}^t}{C_{\text{BA}}}
		\end{eqnarray}
		where $SOC^t$ and $SOC^{t+1}$ denote the charging state of BA at time $t$ and $t+1$. $P_{\text{BA}}$ is the charging-discharging power of BA. Here, we assume $P_{\text{BA}}> 0$ when the BA is discharging, and $P_{\text{BA}}<0$ when the BA is charging. The $\eta_{\text{ch}}$ and $\eta_{\text{dch}}$ are the charging and discharging efficiencies. $\delta$ denotes the discharging rate, which is set as $0.2\%$. $C_{\text{BA}}$ represents the capacity of BA.}
	
	{\color{black}The operation of BA would bring about the costs due to the amortized purchase and maintenance, which is formulated by the following equation \cite{CostModel}:
		\begin{eqnarray}
			\small
			&\begin{split}
				C(P_{\text{BA},j})=&a_{\text{BA},j}(P_{\text{BA},j}+3P_{\text{BA},j}^{\text{max}}(1-SOC))^2\\ 
				&+ b_{\text{BA},j}(P_{\text{BA},j}+3P_{\text{BA},j}^{\text{max}}(1-SOC)) + c_{\text{BA},j}
			\end{split}\\
			&P_{\text{BA},j}^{\text{max}}<P_{\text{BA},j}<P_{\text{BA},j}^{\text{max}}\label{eq:constrain_BA}
	\end{eqnarray}}
	{\color{black}where $C(P_{\text{BA},j})$ represents the cost of the $j$th BA. $a_{\text{BA},j}$, $b_{\text{BA},j}$ and $c_{\text{BA},j}$ are cost coefficients of the $j$th BA. $P_{\text{BA},j}^{\text{max}}$ and $P_{\text{BA},j}^{\text{min}}$ are the upper and lower bounds of BA output power.
		
		\subsubsection{Network Power Loss of MG}
		Practically, there exists the power loss because of the operation of generators and the transmission of energy in the MG. The power loss usually corresponds to the active generation power and can be estimated as follows \cite{powerloss}:}
	{\color{black}\begin{equation}
			\small
			\lambda_{\text{CG}}=\frac{\partial P_{\text{loss}}}{\partial P_{\text{CG}}}, \lambda_{\text{REG}}=\frac{\partial P_{\text{loss}}}{\partial P_{\text{REG}}},\lambda_{\text{BA}}=\frac{\partial P_{\text{loss}}}{\partial P_{\text{BA}}}
		\end{equation}
		where $\lambda_{\text{CG}}$, $\lambda_{\text{REG}}$ and $\lambda_{\text{BA}}$ represent the power loss coefficients of CG, REG and BA, respectively. According to Ref. \cite{powerloss}, $\lambda_{\text{CG}}$, $\lambda_{\text{REG}}$ and $\lambda_{\text{BA}}$ are recommended to be set in $[0.01,0.02]$. Therefore, they are set as 0.02 in this paper.
		
		Then, the power loss $P_{\text{loss}}$ can be given by the following equation \cite{powerloss}:
		\begin{equation}
			\small
			P_{\text{loss}}=\sum_{i=1}^{n_{\text{CG}}}\lambda_{\text{CG}}P_{\text{CG,i}} + \sum_{j=1}^{n_{\text{REG}}}\lambda_{\text{REG}}P_{\text{REG,j}} +\sum_{k=1}^{n_{\text{BA}}}\lambda_{\text{BA}}P_{\text{BA,k}}\label{eq:plossdefine}
		\end{equation}
		where $n_{\text{CG}}$, $n_{\text{REG}}$ and $n_{\text{BA}}$ are the numbers of CGs, REGs and BAs in the isolated MG, respectively.}
	
	\subsection{\textcolor{black}{Isolated MG Energy Management Model via MDP and Physics-Informed Reward}}
	\textcolor{black}{Since the energy management center of the MG is an agent which is trained by the DRL algorithm, the above isolated MG model should be reformulated as the MDP model. In addition, considering the physical feasibility of the agent, the definition of reward is designed to integrate the physical-informed rules, which are presented as follows.}
	
	\subsubsection{State}
	In this paper, we consider a 24-hour scheduling of the MG, and each hour is denoted by $t\in\{1,2,...,24\}$. The state of MG at time $t$ includes the energy operation information, which is defined as follows:
	{\color{black}
		\begin{equation}
			\small
			s_t = \{P_{\text{L}}^{t-1}, P_{\text{REG},1}^{t-1},...,P_{\text{REG},n_{\text{REG}}}^{t-1}, SOC^{t-1}, E_\lambda^{t-1}\}
		\end{equation}
		where $s_t$ indicates the state of MG at time $t$; $P_{\text{L}}^{t-1}$ and $P_{\text{REG,} i}^{t-1}$ stand for the load demand and the $i$th REG at time $t-1$. In addition, the $E_\lambda^{t-1}$ is the electricity price in the transaction between the MG and the distribution power network.
	}
	
	\subsubsection{Action}
	\textcolor{black}{The action $a_t$ is generated by the agent, which controls the power outputs of the CGs and BAs at each time $t$, according to the state $s_t$. In this study, it is defined as follows:}
	\begin{equation}
		\small
		a_t = \{P_{\text{CG},1}^t,...,P_{\text{CG},n_{\text{CG}}}^t,P_{\text{BA},1}^t,...,P_{\text{BA},n_{\text{BA}}}^t\}
	\end{equation}
	
	\textcolor{black}{In DRL, the agent is normally a neural network, which is difficult to produce consistent and feasible in the early training stage. Therefore, the actions are enforced to fulfill the output constraints provided in Eqs (\ref{eq:constrain_CG}) and (\ref{eq:constrain_BA})
		\begin{eqnarray}
			\small
			P_{\text{CG},i}^t = \text{clip}(P_{\text{CG},i}^t,P_{\text{CG},i}^{\text{min}},P_{\text{CG},i}^{\text{max}}),i\in[1,n_{\text{CG}}]\\
			P_{\text{BA},j}^t = \text{clip}(P_{\text{BA},j}^t,P_{\text{BA},j}^{\text{min}},P_{\text{BA},j}^{\text{max}}),j\in[1,n_{\text{BA}}]
		\end{eqnarray} 
		where $\text{clip}(t,t_{\text{min}},t_{\text{max}})$ is the clip function, which returns $t_{\text{max}}$ if $t>t_{\text{max}}$, and $t_{\text{min}}$ if $t<t_{\text{min}}$.}
	
	\subsubsection{Reward}
	The design of the reward significantly impacts the performance of the DRL training. A specific physical task-oriented reward would endow interpretability to the strategy of the agent \cite{psysic_inform}. However, in some classical reinforcement learning tasks, such as CartPole \cite{cart_pole} and Atari Games \cite{Atari}, the design of their rewards is independent of the physical characteristic of the problem. For instance, in CartPole, the reward is set as 0 if the action is available. Such the intuitive design of reward may mislead the agent thus slowing down the training process and decreasing the interpretability of the agent strategy, it is not suitable for the MG energy management. \textcolor{black}{Normally, the MG agent is expected to operate economically while ensuring the self energy-sufficiency. Therefore, considering the physical characteristic of the MG, the reward is designed as physics-informed to satisfy the two explicit targets, i.e., the training of the agent and realizing the requirements of operation cost and self energy-sufficiency, simultaneously. The reward function is defined as follows.}
	
	\begin{equation}
		\small\color{black}
		\begin{split}
			r_t = &-w_{\text{C}}\left(\sum_{i=1}^{n_{\text{CG}}}C(P^t_{\text{CG},i}) + \sum_{j=1}^{n_{\text{BA}}}C(P^t_{\text{BA},j})\right) \\
			&-w_{\text{de}}E_{l}(t)\times \text{abs}(P_{\text{de}}^t)\label{eq:reward}
		\end{split}
	\end{equation}
	\textcolor{black}{where $r_t$ is the reward value at time $t$, and $E_{l}(t) \geq 0$ indicates the price of purchasing electricity from the distribution power grid. $w_{\text{C}}\in[0,1]$ and $w_{\text{de}}\in[0,1]$ indicate the weights to limit the order of magnitude of reward.} $\text{abs}(\cdot)$ stands for the absolute function. $P^t_{\text{de}}$ evaluates the deviation between load demand and real generation, which is formulated by:
	{\color{black}
		\begin{equation}
			\small
			P_{\text{de}}^t = P_{\text{L}}^{t} - \left(\sum_{i=1}^{n_{\text{CG}}}P_{\text{CG,i}}^t+\sum_{j=1}^{n_{\text{REG}}}P_{\text{REG,j}}^t+\sum_{k=1}^{n_{\text{BA}}}P_{\text{BA,k}}^t - P_{\text{loss}}^t\right)\label{eq:reward2}
	\end{equation}}
	
	\textcolor{black}{In this study, the physics-informed reward is composed of two physical targets of MG, i.e., operation costs and self energy-sufficiency. They are formulated as $\left(\sum_{i=1}^{n_{\text{CG}}}C(P^t_{\text{CG},i}) + \sum_{j=1}^{n_{\text{BA}}}C(P^t_{\text{BA},j})\right)$, and $E_{l}(t)\times \text{abs}(P_{\text{de}}^t)$, respectively. To keep the order of magnitude of the reward consistent, the self energy-sufficiency is designed as $P_{\text{de}}^t$ times $E_{l}(t)$.} Since the reward is related to the physical valuables, i.e., $P^t_{\text{CG},i}$, $P^t_{\text{BA},j}$ and $P_{\text{de}}^t$, it can be endowed the physical meaning. In this way, the reward is able to guide the agent to produce a series of actions that minimize the generation costs of CGs and BAs while ensuring self energy-sufficiency.

	\subsection{Decentralized Multi-Microgrid Energy Management Model}
\begin{figure}[bhtp]
		\centering
		\includegraphics[width=3.5in]{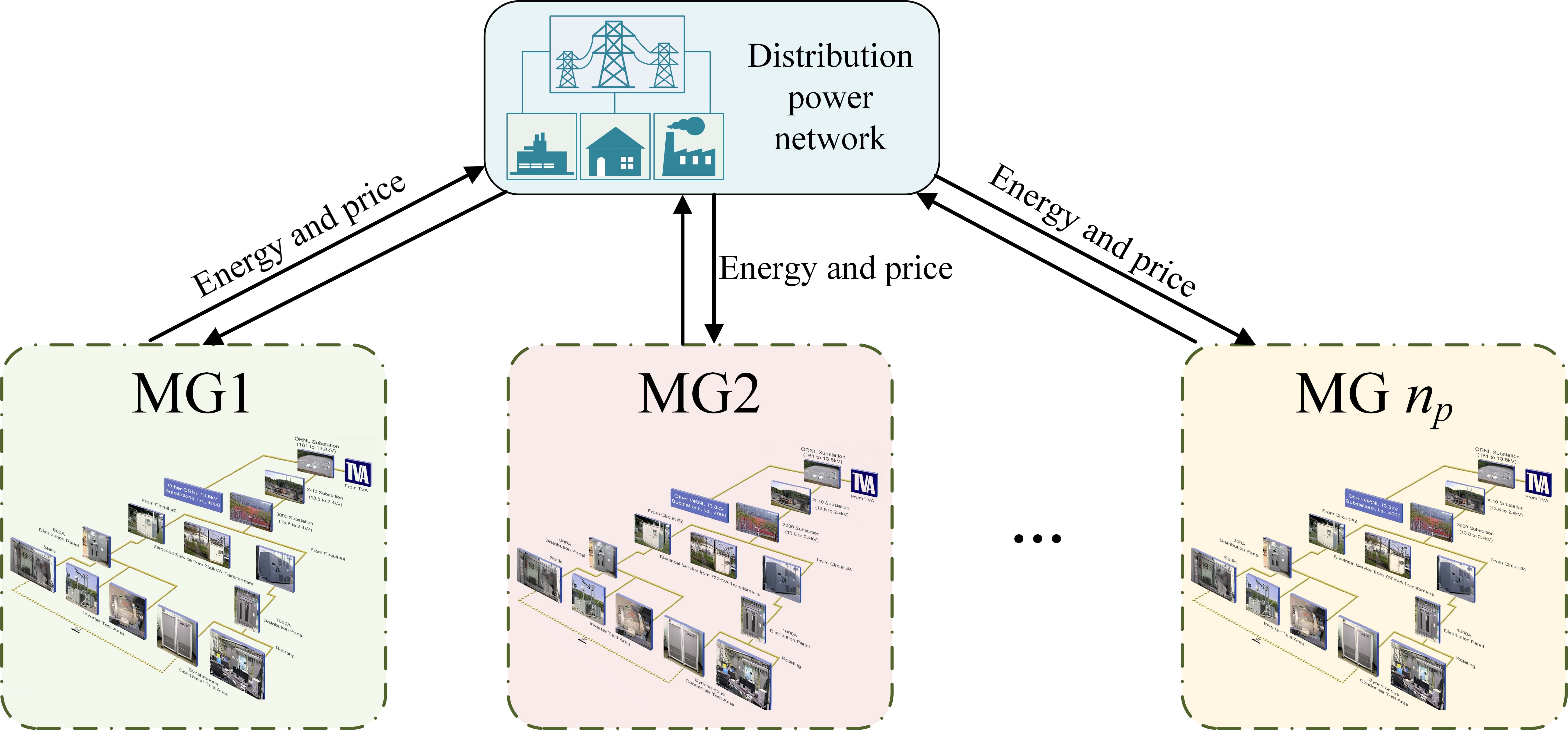}
		\caption{The structure of MMG system.}\label{fig:multiMG}
	\end{figure}

	{\color{black}As shown in Fig. \ref{fig:multiMG}, a decentralized MMG model that contains $n_p$ MGs is considered in this paper. These MGs are connected to the distribution power network, and the energy transaction between MGs is also allowed. Each MG is controlled by an agent, which observes the state $s_t$ of MG and provides the action $a_t$.}
	
	{\color{black}Since the MG is encouraged to maximize the physics-informed reward $r_t$ for achieving energy self-sufficiency and economic operation, the target of the MMG should be the maximum of the systematic rewards $r_{\text{sys},t}$, which can be represented by the sum of rewards obtained by all the MG agents.} The $r_{\text{sys},t}$ is given by
	\begin{equation}
		\small
		\setlength{\abovedisplayskip}{3pt}
		\setlength{\belowdisplayskip}{3pt}
		r_{\text{sys},t} = \sum_{i=1}^{n_p} r^i_t = \sum_{i=1}^{n_p} -\epsilon_i\times \text{abs}(P_{i,\text{de}}^t)
	\end{equation}
	where $r_t^i$ represents the reward obtained by the $i$th MG agent at time $t$. $\epsilon_i$ and $P_{i,\text{de}}^t$ are the shrinkage coefficient and deviation of MG $i$.
	
	Besides, since the load demand of MG cannot be known in advance, excessive or insufficient power generation of an isolated MG is unavoidable, thus the energy transaction in the MMG system is inevitable.
	Therefore, the energy transaction mechanism between different MGs is developed, which is given below.
	
	That is, MG is allowed to conduct energy transactions with the distribution power network and other MGs, as shown in Fig. \ref{fig:multiMG}. If the generated power of MG $i$ exceeds its load demand at time $t$, the excess energy will be sold to other MGs with a price $E_i(t)$. If the demand of MG $i$ cannot be satisfied, the MG will purchase electricity from MG $j$, \textcolor{black}{which has the lowest price of the whole participated MGs.}
	\begin{equation}
		\small
		j=\mathop{\arg\min}\limits_{l} E_{l}(t)\times L_l,l\in[1,2,...,n_p]
	\end{equation}
	where $L_l$ indicates whether the generation of MG $l$ exceeds its demand. The $L_l$ is set as $1$ if the demand is exceeded or set as infinite if not. The MGs will preferentially purchase the surplus power generated by other MGs. When the MG generations are fully consumed, the distribution power network will provide power with price $E_{\text{dpn}}(t)$, which is usually higher than $E_{l}(t),l\in[1,2,...,n_p]$.
	
	\section{Federated Multi-Agent Deep Reinforcement Learning Algorithm}
	As discussed above, the MMG is decentralized, thus each MG agent has a high autonomy. However, the decentralized structure threatens the generalization performance of the agent, because the diversity of the data in isolated MG is limited, which may make the agent getting trap into a local optima. To tackle this issue, we propose a federated multi-agent deep reinforcement learning (F-MADRL) algorithm. The FL is used to improve the generalization of the agent during training while ensuring data privacy. 
	
	There are two characteristics in the FL, one is called participant and the other is termed as the collaborator. The participant $j,j\in[1,n_p]$, is denoted as a neural network model $f^j_{w_j}$. It conducts self-training at the local and uploads its parameters $\bm{w_j}$ to the collaborator periodically, where $n_p$ is the number of participants which are processed in parallel. Constrained by the data privacy, the participant $f^j_{w_j}$ only trains on the local dataset, which may cause the insufficient training since the capacity and diversity of the data are limited. The FL could tackle this problem through the following steps. \textcolor{black}{First, at the training epoch $e,e\in[1,N_e]$, the model of $j$th participant is defined as $f^j_{w_j^e}$, which conducts self-training to obtain the parameters $\bm{w_j^e}$, where $N_e$ is the total number of training epoches.} Then, each participant uploads its parameters to the collaborator and constructs a parameter list $\bm{w_e}=\left[\bm{w_1^e},\bm{w_2^e},...,\bm{w_{n_p}^e}\right]$. The collaborator calculates the weight average of $\bm{w_e}$ to estimate a global model $f_G^{e+1}$ with parameters $\bm{\overline{w}_G^{e+1}}$. After aggregation, the collaborator broadcasts $\bm{\overline{w}_G^{e+1}}$ to all the participants and replaces their own parameters, i.e., $\bm{\overline{w}_G^{e+1}}=\bm{w_1^{e+1}}=\bm{w_2^{e+1}}...=\bm{w_{n_p}^{e+1}}$. The aggregating mechanism of FL is formulated by the following equations:
	\begin{eqnarray}
		\small
		\bm{w_j^{e+1}}= \bm{\overline{w}^{e}_G}-\eta \nabla F_j(\bm{w_j^{e}}), \forall j\\
		\bm{\overline{w}_G^{e+1}}= \sum_{j=1}^{n_p}\frac{1}{n_p}\bm{w_j^{e+1}}\label{eq:aggregate}
	\end{eqnarray}
	where $\eta$ and $F_j(\cdot)$ are the learning rate and local loss function of the $j$th participant, respectively.
	
	In this paper, the participant can be considered as the agent in each MG and the collaborator is a server that takes the responsibility for aggregating and broadcasting the parameters. The F-MADRL aims to solve the following distributed optimization model:
	\begin{equation}
		\small
		\setlength{\abovedisplayskip}{3pt}
		\setlength{\belowdisplayskip}{3pt}
		\min_{\bm{\overline{w}_G^{e}}}F(\bm{\overline{w}_G^{e}})=\sum_{j=1}^{n_p}p_j F_j(\bm{w_j^e})\label{eq:global_loss}
	\end{equation}	
	\textcolor{black}{where $F(\cdot)$ is the global loss function. $p_j $ represents the relative weight of each MG agent on the global model, and $p_j>0$, $\sum_{k=1}^{n_p}p_j = 1$. We set $p_j=|D_j|/\sum_{j=1}^{n_p}|D_j|$, where $|D_j|$ is the data size used for the local training of $j$th MG. Note that $F(\cdot)$ cannot be directly computed without the information sharing of each participant.}
	
\begin{figure}[htp]
		\centering
		\includegraphics[width=3.5in]{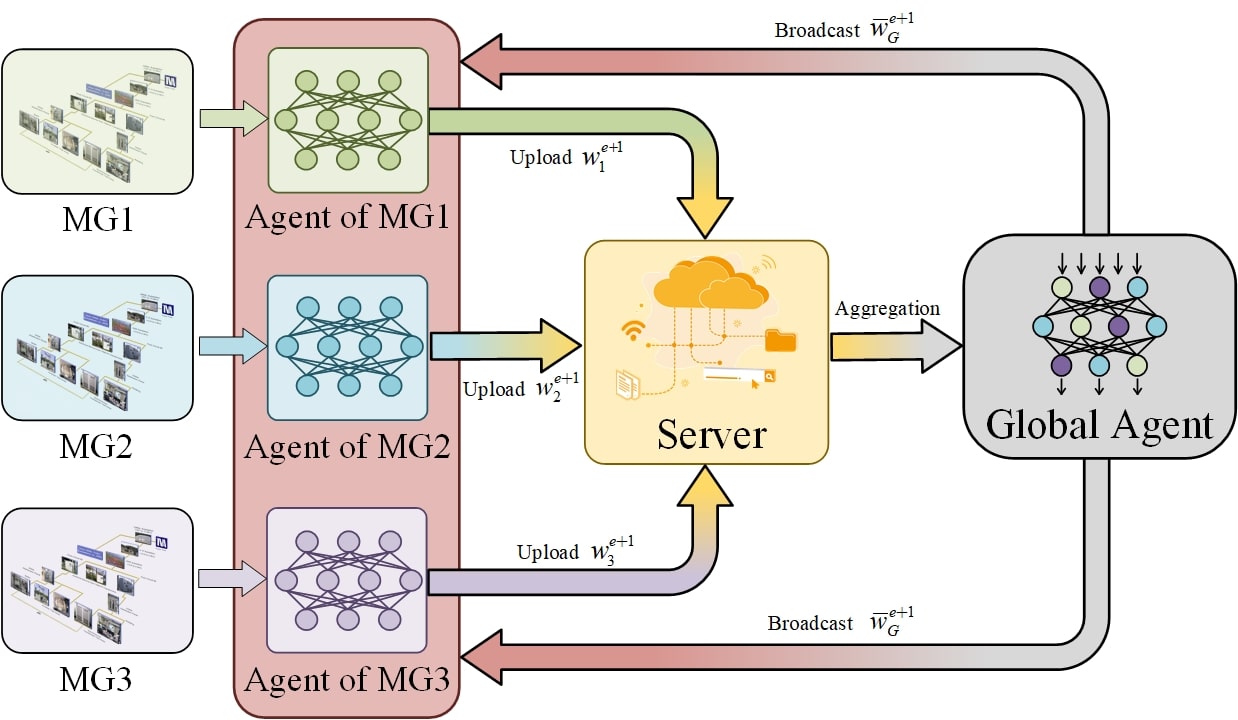}
		\caption{\color{black}The proposed federated multi-agent deep reinforcement learning algorithm.}\label{fig:FLStructure}
	\end{figure}

	The overall structure of F-MADRL is illustrated in Fig. \ref{fig:FLStructure}. At the epoch $e$, the agent in three MGs are firstly replaced by the global agent in the $(e-1)$th epoch. Then, the three MG agents conduce self-training to obtain parameters, which are uploaded to the server for aggregation. Next, the global agent would be built on the server, and the parameters will be broadcasted to the MG agents for the $(e+1)$th epoch.
	
	It can be learned from the figure that the F-MADRL contains two parts. \textcolor{black}{One is executed on server, which can be considered as the collaborator and the other is executed on MG agents, can be considered as the participant. The procedures running on the server and MG agent are provided in the following subsections, respectively.}
	
	\subsection{The F-MADRL: Server Part}
	The F-MADRL proceed on the server mainly focus on the aggregating and broadcasting of the agent parameters, and its procedure is shown in \textbf{Algorithm 1}.
	
	At the beginning of the training epoch of F-MADRL, the server would build a global agent with the parameter $\bm{\overline{w}_G^{0}}$, which is then broadcasted to each MG agent for self-training. Since the agents update their parameters in parallel, the server aggregates the parameters list $\bm{w_e}=[\bm{w_1^e},\bm{w_2^e},...,\bm{w_{n_p}^e}]$ by Eq. (\ref{eq:aggregate}). Furthermore, the aggregated parameters $\bm{\overline{w}_G^{e+1}}$ are used to update the global model parameters and broadcast to the MG agents for the training of epoch $e+1$.
	\begin{breakablealgorithm}
		\caption{The federated multi-agent deep reinforcement learning algorithm on the server.}
		\begin{algorithmic}[1]	
			\State\textbf{Execute on the server:}
			\State Initialize the model parameters $\bm{\overline{w}_G^{0}}$ and broadcast them to the MG agents.
			\For{Global epoch $e=1$ to $N_e$}
			\For{MG agent $j=1$ to $n_p$ \textbf{\textcolor{black}{parallelly}}}
			\State Update the MG parameter $\bm{w_j^{e}}$ at the local agent.
			\State Store the $\bm{w_j^{e}}$.
			\State Upload the $\bm{w_j^{e}}$ to the server
			\EndFor
			\State Receive the parameters from each MG agent and construct
			\State $\bm{w_e}\leftarrow[\bm{w_1^e},\bm{w_2^e},...,\bm{w_{n_p}^e}]$
			\State Aggregating the model parameters through 
			\State $\bm{\overline{w}_G^{e+1}}= \sum_{j=1}^{n_p}\frac{1}{n_p}\bm{w_j^{e+1}}$.
			\State Broadcast the $\bm{\overline{w}_G^{e+1}}$ to other MG agents.
			\EndFor
		\end{algorithmic}	
	\end{breakablealgorithm}

	\vspace{-0.3cm}
	
	\subsection{The F-MADRL: MG Agent}
	On the other hand, the MG agent adopts the self-training in the procedure of F-MADRL and cooperates with the server. When the MG agents receive the parameter $\bm{\overline{w}_G^{e}}$ from global model at the epoch $e$, their parameters are replaced by $\bm{\overline{w}_G^e}$, i.e., $\bm{w_j^e} =\bm{\overline{w}_G^e}$. Then, each MG agent executes $N_i$ individual self-training epochs in parallel. Afterwards, the parameters of the MG agent at the last self-training epoch, namely $\theta_{N_i},\mu_{N_i}$ are stored and uploaded to the server. 
	
	In this paper, each MG agent performs self-training with a famous deep reinforcement learning algorithm, namely PPO, to obtain the optimal policy $\pi$. There are two types of deep neural networks, namely, actor and critic, defined by the MG agent. Actor $\pi^\theta$ is parameterized by $\theta$, which aims to produce the action, and the critic is denoted as $V^\mu$, which is parameterized by $\mu$.
	\begin{figure}[hbtp]
		\centering
		\includegraphics[width=3.2in]{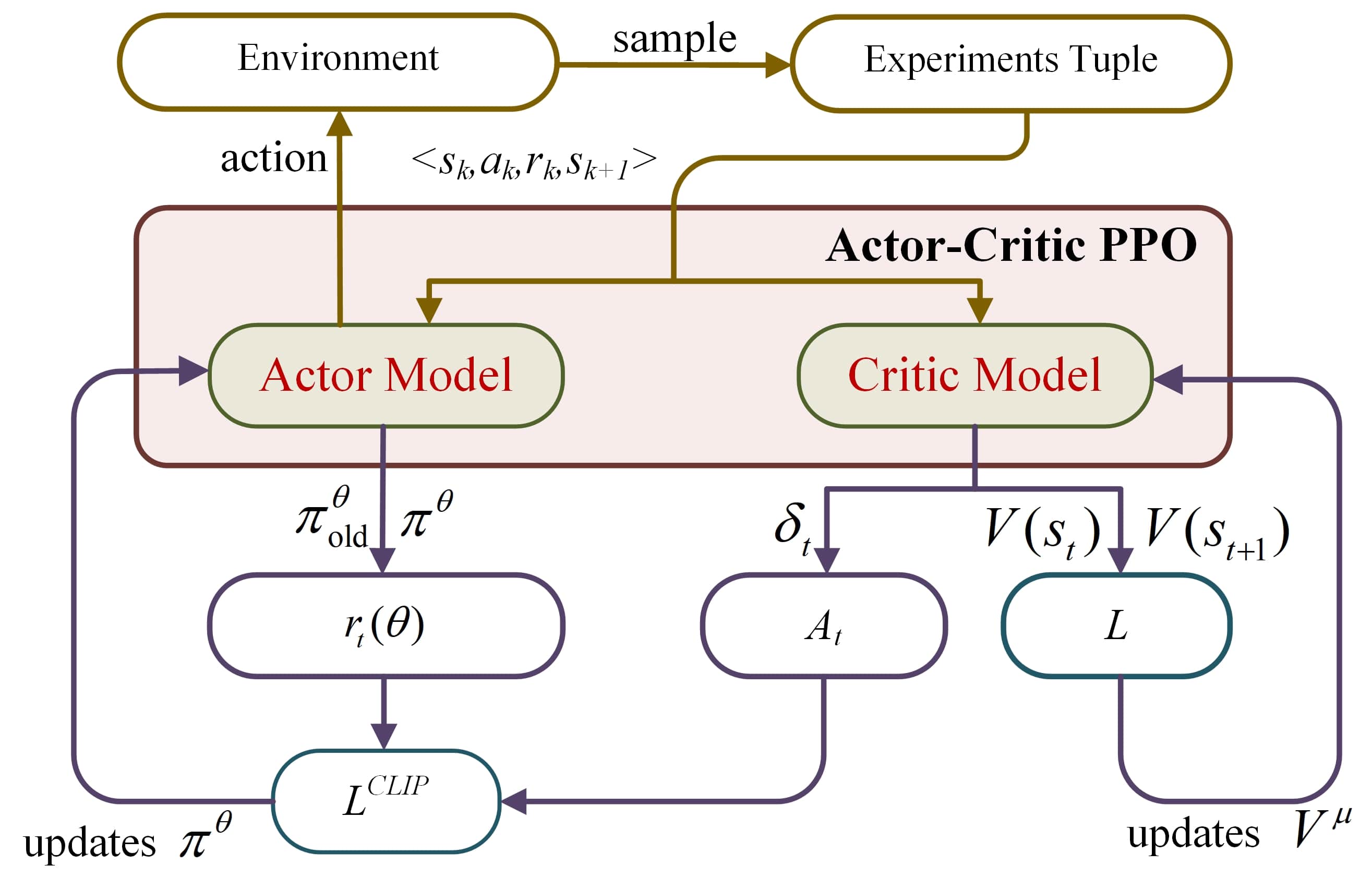}
		\caption{\textcolor{black}{The procedure of the self-training of MG agent.}}\label{fig:PPOStructure}
	\end{figure}
	
	The overall training process of the self-training procedure during one episode is illustrated in Fig. \ref{fig:PPOStructure}. First of all, the experiment tuples $T$ are sampled $T = \{\langle s_0,a_0,r_0,s_{1}\rangle, \langle s_1,a_1,r_1,s_{2}\rangle,...,\langle s_{U},a_{U},r_{U},s_{U+1}\rangle\}$ where $U$ indicates the length of $T$. Then, the loss function of the actor at $k$th episodes is calculated, which is defined as follows:
	\begin{equation}
		\small
		\begin{split}
			\mathcal{L}^{\text{C}}=\mathbb{E}_{s,a \sim T}\bigg[\min (\frac{\pi^\theta_k(a \mid s)}{\pi^\theta_{k-1}(a \mid s)} A^{\pi^\theta_{{k}}}_{s, a},\\
			\text{clip}(\frac{\pi^\theta_k(a \mid s)}{\pi^\theta_{k-1}(a \mid s)},1-\epsilon,1+\epsilon) A^{\pi^\theta_{{k}}}_{s, a})\bigg]\\
		\end{split}\label{eq:loss_actor}
	\end{equation}
	where $\mathbb{E}_{s,a \sim T}[\cdot]$ represents the empirical average over the sampled experiment tuples $T$. The $\pi_{k-1}$ and $\pi_k$ stand for the previous and new policy, respectively. The $\epsilon$ is the clip parameter. {\color{black}
		$A^{\pi_{{k}}}_{s, a}$ stands for the advantage, which measures if the action is worth taking by comparing the action value and state value:
		\begin{equation}
			A^{\pi_{{k}}}_{s_t, a_t}=E_{\tau}[U|s_0=s_t,a_0=a_t]-V(s_t)=Q(s_t,a_t)-V(s_t)\label{eq:R3-2_defined_A}
		\end{equation}
		
		However, the $A^{\pi_{{k}}}_{s, a}$ cannot be directly obtained since $Q(s_t,a_t)$ is difficult to be determined. In this way, the generalized advantage estimation method is implemented in this study:
		\begin{equation}
			A^{\pi^\theta_{{k}}}_{s, a} = \delta_0^V +(\gamma \lambda)\delta_{1}^V+(\gamma \lambda)^2\delta_{2}^V,...,+(\gamma \lambda)^{U-t+1}\delta_{U-1}^V
		\end{equation}
		where $\gamma \in[0,1]$ and $\lambda \in[0,1]$ represent the discount factor and a hyperparameter that adjusts the tradeoff between bias and variance of the estimation. Note that the variance would be increased when raising $\lambda$ while the bias is decreased accordingly. According to the recommendation of Ref. \cite{PPO}, $\lambda$ is set as 0.95. $\delta_k^V$ is calculated by:
		\begin{equation}
			\delta_k^V = r_k+\gamma V^\mu_k(s_{t+1})-V^\mu_k(s_t)
		\end{equation}
	}
	where $V^\mu_k(s_{t+1})$ and $V^\mu_k(s_t)$ are given by the critic, which is trained by the loss function $\mathcal{L}^V$:
	\begin{equation}
		\small 
		\mathcal{L}^V=\mathbb{E}_{s,a \sim T}\left[(\gamma V^\mu_k(s_{t+1}) + r(s_t, a_t) -V^\mu_k(s_t))^2\right].\label{eq:loss_critic}
	\end{equation}
	
	With the above equations, the parameter $\theta$ and $\mu$ of the actor and the critic can be updated by the following equations:
	\begin{eqnarray}
		\small
		&\theta_{k+1} = \theta_{k}+\eta_{\pi}\nabla_{\theta_{k}}\mathcal{L}^{\text{C}}\\
		&\mu_{k+1} = \mu_{k}+\eta_{V}\nabla_{\mu_{k}}\mathcal{L}^V
	\end{eqnarray}
	where $\eta_\pi$ and $\eta_V$ are the learning rates of actor and critic. \textcolor{black}{Since both $\mathcal{L}^{\text{C}}$ and $\mathcal{L}^V$ are optimized in each MG agent of the proposed F-MADRL algorithm, they are the local loss functions which construct the global loss following Eq. (\ref{eq:global_loss}).}

	Overall, the F-MADRL algorithm applied on the server can be summarised in the \textbf{Algorithm 2}.

	\subsection{The Theoretical Convergence Analysis of the F-MADRL}
	In this section, the convergence of the F-MADRL is evaluated. At first, the following assumptions considering the function $F_k,k\in[1,n_p]$ are made, by referring to Ref. \cite{Proof}.
	
	{\color{black}
		\textbf{Assumption 1:} The $F_k$ is $L$-smooth, $\forall \bm{w}, \bm{w'}$, $\Vert \Delta F_k(w)-\Delta F_k(w')\Vert_2\leq L\Vert w-w'\Vert$; $F_k(\bm{w})\leq F_k(\bm{w'})+\nabla F_k(\bm{w'})^T(\bm{w}-\bm{w'})+\frac{L}{2}\|\bm{w'}-\bm{w}\|^2_2$.}
	
	\begin{breakablealgorithm}
		\caption{The federated multi-agent deep reinforcement learning algorithm on the MG agent.}
		\begin{algorithmic}[1]	
			\State\textbf{Execute on each MG agent:}
			\State Parallel running on $j, j\in[1,n_p]$ agent at global epoch $e$
			\State Receive the parameters from the server $\bm{w_j^e} \leftarrow\bm{\overline{w}_G^e}$
			\For{Individual training epoch $i=1$ to $N_i$}
			\State \textcolor{black}{Collect the experience tuple $T = \{\langle s_0,a_0,r_0,s_{1}\rangle,$ $ \langle s_1,a_1,r_1,s_{2}\rangle,...,\langle s_{U},a_{U},r_{U},s_{U+1}\rangle\}$}
			\State Compute the discounted factor:
			\State $\delta_k^V \leftarrow r_k+\gamma V^\mu_k(s_{t+1})-V^\mu_k(s_t)$
			\State Estimate the advantage:
			\State $A^{\pi^\theta_{{k}}}_{s, a} \leftarrow \delta_0^V +(\gamma \lambda)\delta_{1}^V+(\gamma \lambda)^2\delta_{2}^V,...,+(\gamma \lambda)^{U-t+1}\delta_{U-1}^V$
			\State Calculate the loss function of the actor:
			\State $\mathcal{L}^{\text{C}}\leftarrow \mathbb{E}_{s,a \sim T}\bigg[\min (\frac{\pi^\theta_i(a \mid s)}{\pi^\theta_{i-1}(a \mid s)} A^{\pi^\theta_{{i}}}_{s, a},\text{clip}(\frac{\pi^\theta_i(a \mid s)}{\pi^\theta_{i-1}(a \mid s)},1-$
			\State $\epsilon,1+\epsilon)A^{\pi^\theta_{{i}}}_{s, a})\bigg]$
			\State Update the actor parameter:
			\State $\theta_{i+1} \leftarrow \theta_{i}+\eta_{\pi}\nabla_{s,a \sim T}\mathcal{L}^{\text{C}}$
			\State Calculate the loss function of the critic:
			\State $\mathcal{L}^V\leftarrow \mathbb{E}_{s,a \sim T}\left[(\gamma V^\mu_i(s_{t+1}) + r(s_t, a_t) -V^\mu_i(s_t))^2\right]$
			\State Update the critic parameter:
			\State $\mu_{i+1} \leftarrow \mu_{i}+\eta_{V}\nabla_{\mu_{i}}L^V$
			\EndFor
			\State Store the network parameters. $\bm{w_j^e}\leftarrow\{\theta_{N_i},\mu_{N_i}\}$
			\State Upload the parameters $\bm{w_j^e}$ to the server.
		\end{algorithmic}	
	\end{breakablealgorithm}
	
	{\color{black}
		\textbf{Assumption 2:} The $F_k$ is $\mu$-strongly convex,  $\forall \bm{w}$,  $\bm{w'}$,  $F_k(w)$$ - \frac{\mu}{2}\Vert F_k \Vert^2 \text{ is convex}$; $F_k(\bm{w})\geq F_k(\bm{w'})+\nabla F_k(\bm{w'})^{T}(\bm{w}-\bm{w'})+\frac{\mu}{2}\|\bm{w'}-\bm{w}\|^2_2$.
	}
	
	Based on the above \textbf{Assumptions}, we have the following \textbf{Lemmas}.
	
	\textbf{Lemma 1: } $F$ is $\mu$-strongly convex and $L$-smooth.
	
	\textbf{Proof: }Straightforwardly from \textbf{Assumption 1} and \textbf{Assumption 2}, in line with the definition of convex, $F$ is the finite-sum of the $F_k$, thus it is $\mu$-strongly convex and $L$-smooth as well.
	
	\textbf{Lemma 2: }$\forall \bm{w}, \bm{w'}\in \mathbb{R}^n$ and $\bm{w_t}=\bm{w}+t(\bm{w}-\bm{w'})$ for $t\in[0,1]$. Then,
	\begin{equation}
		\small
		F(\bm{w})-F(\bm{w'}) = \int_0^1\nabla F(\bm{w_t})^T(\bm{w'}-\bm{w})dt \label{eq:lamma2_0}
	\end{equation}
	and
	\begin{equation}
		\small
		\begin{split}
			F(\bm{w})-F(\bm{w'})-\nabla F(\bm{w})^T(\bm{w'}-\bm{w})\\
			=\int_0^1(\nabla F(\bm{w_t})-\nabla F(\bm{w}))^T (\bm{w'}-\bm{w})dt \label{eq:lamma2_1}
		\end{split}
	\end{equation}
	
	\textbf{Proof: } Eq. (\ref{eq:lamma2_0}) follows the fundamental theorem of calculus. Eq. (\ref{eq:lamma2_1}) follows from Eq. (\ref{eq:lamma2_0}) by subtracting $\nabla F(\bm{w})^T(\bm{w'}-\bm{w})$ from both sides of the equation.
	
	\textbf{Lemma 3: }If $F$ is smooth and $\mu$-strongly convex for $\mu>0$, then for the $\bm{w_*}=\mathop{\arg\min}\limits_{\bm{w}}F(\bm{w})$,
	
	\begin{equation}
		\small
		\frac{1}{2\mu}\|\nabla F(\bm{w})\|^2_2 \geq F(\bm{w})-\bm{w_*} \geq \frac{\mu}{2}\|\bm{w}-\bm{w_*}\|^2_2
	\end{equation}
	
	\textbf{Proof: } Applying the \textbf{Lemma 2}, we have
	\begin{equation}
		\small
		F(\bm{w})\geq F(\bm{w_*})+\nabla F(\bm{w_*})^T(\bm{w}-\bm{w_*})+\frac{\mu}{2}\|\bm{w_*}-\bm{w}\|^2_2 \label{ineq:convex}
	\end{equation}
	
	Using the fact that $\nabla F(\bm{w_*})=0$, we yield
	\begin{equation}
		\small
		F(\bm{w})-F(\bm{w_*})\geq \frac{\mu}{2}\|\bm{w_*}-\bm{w}\|^2_2
	\end{equation}
	which is the right side of (\ref{ineq:convex}).
	
	Note that
	\begin{equation}
		\small
		F(\bm{w_*})\geq \min_{\bm{y}} F(\bm{y}) +\nabla F(\bm{w})^T(\bm{y}-\bm{w})+\frac{\mu}{2}\|\bm{w}-\bm{y}\|^2_2
	\end{equation}
	since $\bm{y}=\bm{w}-\frac{1}{\mu}\|\nabla\|F(\bm{w})$ minimizes the right side of the above inequality, we yield
	\begin{equation}
		\small
		\begin{split}
			&\min_{\bm{y}} \left[ F(\bm{y}) +\nabla F(\bm{w})^T(\bm{y}-\bm{w})+\frac{\mu}{2}\|\bm{w}-\bm{y}\|^2_2\right ]\\
			&\geq F(\bm{w})-\frac{1}{2\mu}\|\nabla F(\bm{w})\|^2_2
		\end{split}
	\end{equation}
	namely,
	\begin{equation}
		\small
		\frac{1}{2\mu}\|\nabla F(\bm{w})\|^2_2\geq F(\bm{w})-F(\bm{w_*})
	\end{equation}
	
	\textbf{Theorem 1: } \textcolor{black}{Considering the $F$ is $L$-smooth and $\mu$-strongly convex, let $\bm{w_*}=\mathop{\arg\min}\limits_{\bm{w}} F(w)$, and $w$ is the parameter at the $k$th iteration. } Then,
	\begin{equation}
		\small
		F(\bm{w_{k}})-F(\bm{w_*})\leq \left( 1-\frac{\mu}{L} \right)^k(F(\bm{w_0})-F(\bm{w_*}))
	\end{equation}
	Consequently, it requires $\frac{L}{\mu}\log(\frac{F(\bm{w_0})-F(\bm{w_*})}{\epsilon})$ iterations to find $\epsilon$-optimal solution.
	
	\textbf{Proof: } Applying \textbf{Lemma 2}, we have
	\begin{equation}
		\small
		\begin{split}
			&|F(\bm{w})-F(\bm{w'})-\nabla F(\bm{w})^T(\bm{w'}-\bm{w})|\\
			&\leq |\int_0^1(\nabla F(\bm{w_t})-\nabla F(\bm{w}))^T (\bm{w'}-\bm{w})dt|\\
			&\leq \int_0^1(\nabla \|F(\bm{w_t})-\nabla F(\bm{w})\|) \|\bm{w'}-\bm{w}\|dt
		\end{split}
	\end{equation}
	
	Based on the \textbf{Assumption 1}, $\|F(\bm{w_t})-\nabla F(\bm{w}))\|\leq t\|\bm{w}-\bm{w'}\|$. Note that $\bm{w_t}-\bm{w} = t(\bm{w}-\bm{w'})$. Then, we have
	\begin{equation}
		\small
		\begin{split}
			&|F(\bm{w})-F(\bm{w'})-\nabla F(\bm{w})^T(\bm{w'}-\bm{w})|\\
			&\leq \int_0^1(\nabla \|F(\bm{w_t})-\nabla F(\bm{w})\|) \|\bm{w'}-\bm{w}\|dt\\
			&\leq \frac{L}{2}\|\bm{w}-\bm{w'}\|^2_2
		\end{split}
	\end{equation}
	
	Note that
	\begin{equation}
		\bm{w_{k+1}} = \bm{w_{k}} - \eta \nabla F(\bm{w_{k}})
	\end{equation}
	where $\eta>0$ represents the learning rate. Then, we yield
	\begin{equation}
		\small
		F(\bm{w_{k+1}})-(F(\bm{w_k})+\eta\|\nabla F(\bm{w_k})\|^2_2)\leq \frac{\eta^2 L}{2}\|\nabla F(\bm{w_k})\|^2_2
	\end{equation}
	if we pick $\eta=\frac{1}{L}$, then $F(\bm{w_{k+1}})\leq F(\bm{w_k})-\frac{1}{2L}\|\nabla F(\bm{w_k})\|^2_2$.
	From \textbf{Lemma 3}, $\|\nabla F(\bm{w})\|^2_2\geq 2\mu (F(\bm{w})-F(\bm{w_*}))$. When putting them together,
	\begin{eqnarray}
		\color{black}
		\small
		\begin{split}
			F(\bm{w_{k+1}})-F(\bm{w_*})&\leq F(\bm{w_k})-F(\bm{w_*})-\frac{1}{2L}\|\nabla F(\bm{w_k})\|^2_2\\
			&\leq F(\bm{w_k})-F(\bm{w_*})\\
			&\quad-\frac{\mu}{L}(F(\bm{w_k})-F(\bm{w_*}))\\
			&=(1-\frac{\mu}{L})(F(\bm{w_k})-F(\bm{w_*}))
		\end{split}
	\end{eqnarray}
	
	Repeatedly applying this bound yields
	\begin{eqnarray}
		\small
		F(\bm{w_{k}})-F(\bm{w_*})\leq \left( 1-\frac{\mu}{L} \right)^k(F(\bm{w_0})-F(\bm{w_*}))
	\end{eqnarray}
	
	Using the fact that $1+x \leq e^x$, the convergence rate is given by picking $k\geq \frac{L}{\mu} \log(\frac{F(\bm{w_0})-F(\bm{w_*})}{\epsilon})$, where $\epsilon$ = $F(\bm{w_{k}})-F(\bm{w_*})$, denoting the error between the loss at epoch $k$ and the optimal one.
	
	\section{Case Study}

		\subsection{Experiment Setup}
		In this section, we conduct case studies based on the modified Oak Ridge National Laboratory Distributed Energy Control Communication lab microgrid test system to demonstrate the effectiveness of the proposed F-MADRL algorithm. Without loss of generality, three MGs are applied to form the MMG system and the parameters of elements in each MG are provided in Table.\ref{tab:paraMG}. A wind turbine and a PV panel are set as the REGs and the corresponding power data is referred from Ref.\cite{WindSolarLoadPrice}. The forecast errors of wind and PV power is assumed to be independent of Gaussian distribution with a 15\% standard variation \cite{WindSolarLoadPrice}. In addition, the time horizon of the experiment is set as 
		24-hour schedule, and the time interval is set to be 1h. \textcolor{black}{The forecast total load demands and the day-ahead market price of each MG are provided in the Table. \ref{tab:predpara} and the forecast error of the load is assumed to follow the Gaussian distribution with a 3\% standard variation \cite{Load2}\cite{Load3}.} It can be learned from the load data that the power demands of MG1 surpass its maximum capacity, which means the MG1 operates in the energy self-insufficient state. On the contrary, the MG2 and MG3 operate in the energy self-sufficient state.

		As for the F-MADRL, the training epoch is set as 1500. \textcolor{black}{Moreover, $\gamma$ and $\lambda$ are set as $0.99$ and $0.95$, receptively \cite{sutton2018reinforcement}.} A famous neural network optimizer, Adam \cite{PPO}, is used to update the F-MADRL, and the learning rate of actor $\eta_\pi$ and critic $\eta_V$ are set as 0.0001 and 0.001. Note that all simulation studies are conducted using Python 3.6.8 with PyTorch 1.7.1.
		{\color{black}
			\begin{table}[htbp]
				\centering
				\caption{The Parameters Setting of Each MG}
				\scalebox{0.85}{
					\begin{tabular}{ccccccc}
						\toprule
						& \multicolumn{1}{c}{} & \multicolumn{1}{p{4.04em}}{$a(\$\text{/kW})$} & \multicolumn{1}{p{4.04em}}{$b(\$\text{/kW})$} & \multicolumn{1}{p{4.04em}}{$c(\$\text{/kW})$} & \multicolumn{1}{p{4.04em}}{$P_{\text{min}}(\text{kW})$} & \multicolumn{1}{p{4.04em}}{$P_{\text{max}}(\text{kW})$} \\
						\midrule
						\multirow{2}[1]{*}{MG1}  & CG    & 0.0081 & 5.72  & 63    & 0     & 200 \\
						& BA    &0.0153 &5.54  & 26    & -50    & 50  \\
						\multirow{2}[0]{*}{MG2}  & CG    & 0.0076 & 5.68  & 365   & 0     & 280 \\
						& BA    &0.0163 &5.64  & 32    & -50    & 50  \\
						\multirow{2}[1]{*}{MG3}  & CG    & 0.0095 & 5.81  & 108   & 0     & 200 \\
						& BA    &0.0173 &5.74  & 38    & -50    & 50  \\
						\bottomrule
					\end{tabular}\label{tab:paraMG}}
			\end{table}%
		}
		{\begin{table*}[htbp]
				\color{black}
				\centering
				\caption{\textcolor{black}{The forecasting wind and PV power, trading price and loads of the three MGs.}}
				\scalebox{0.9}{
					\begin{tabular}{ccccccccccccc}
						\toprule
						Hour  & 1     & 2     & 3     & 4     & 5     & 6     & 7     & 8     & 9     & 10    & 11    & 12 \\
						\midrule
						Wind power (kW) & 51.48  & 38.37  & 43.56  & 40.75  & 27.74  & 30.15  & 28.65  & 23.38  & 21.75  & 34.82  & 27.17  & 30.20  \\
						PV outputs (kW) & 0.00  & 0.00  & 0.00  & 0.00  & 0.00  & 0.00  & 0.16  & 1.77  & 5.30  & 11.60  & 36.64  & 42.68  \\
						\multicolumn{1}{m{15em}}{\centering Trading price among MG and distribution network (\$/kW)} & 8.65  & 8.11  & 8.25  & 8.10  & 8.14  & 8.13  & 8.34  & 9.35  & 12.00  & 9.19  & 12.30  & 20.70  \\
						\multicolumn{1}{m{15em}}{Trading price among MGs (\$/kW)} & 4.33  & 4.06  & 4.13  & 4.05  & 4.07  & 4.07  & 4.17  & 4.68  & 6.00  & 4.60  & 6.15  & 10.35  \\
						Load of MG1 (kW) & 457.70  & 336.50  & 274.90  & 272.60  & 245.30  & 233.70  & 274.60  & 291.00  & 315.70  & 362.40  & 320.00  & 350.00  \\
						Load of MG2 (kW) & 110.50  & 109.85  & 112.45  & 110.50  & 113.75  & 120.25  & 130.00  & 157.95  & 165.10  & 169.00  & 173.55  & 168.35  \\
						Load of MG3 (kW) & 124.71  & 123.98  & 126.91  & 124.71  & 128.38  & 135.43  & 146.72  & 178.26  & 186.33  & 190.73  & 195.87  & 190.00  \\
						\midrule
						&       &       &       &       &       &       &       &       &       &       &       &  \\
						\midrule
						Hour  & 13    & 14    & 15    & 16    & 17    & 18    & 19    & 20    & 21    & 22    & 23    & 24 \\
						\midrule
						Wind power (kW) & 23.52  & 39.48  & 35.74  & 18.06  & 24.27  & 26.26  & 26.77  & 26.22  & 32.84  & 36.02  & 37.23  & 44.12  \\
						PV outputs (kW) & 35.22  & 35.46  & 34.83  & 23.62  & 14.18  & 4.67  & 0.18  & 0.00  & 0.00  & 0.00  & 0.00  & 0.00  \\
						\multicolumn{1}{p{15em}}{\centering Trading price among MG and distribution network (\$/kW)} & 26.82  & 27.35  & 13.81  & 17.31  & 16.42  & 9.83  & 8.63  & 8.87  & 8.35  & 16.44  & 16.19  & 8.87  \\
						\multicolumn{1}{p{15em}}{Trading price among MGs (\$/kW)} & 13.41  & 13.68  & 6.91  & 8.66  & 8.21  & 4.92  & 4.32  & 4.44  & 4.18  & 8.22  & 8.10  & 4.44  \\
						Load of MG1 (kW) & 345.20  & 320.60  & 333.20  & 316.80  & 291.30  & 413.80  & 539.80  & 557.20  & 557.10  & 535.00  & 437.80  & 447.30  \\
						Load of MG2 (kW) & 168.35  & 165.75  & 170.30  & 172.25  & 165.75  & 164.25  & 162.50  & 165.75  & 169.00  & 161.20  & 148.00  & 119.60  \\
						Load of MG3 (kW) & 190.00  & 187.07  & 192.20  & 194.40  & 187.07  & 185.60  & 183.40  & 187.07  & 190.73  & 181.93  & 161.39  & 134.98  \\
						\bottomrule
					\end{tabular}\label{tab:predpara}}
			\end{table*}%
		}
		
		\subsection{Analysis of the F-MADRL algorithm}
		
\begin{figure}[bhpt]
			\centering
			\includegraphics[width=3.5in]{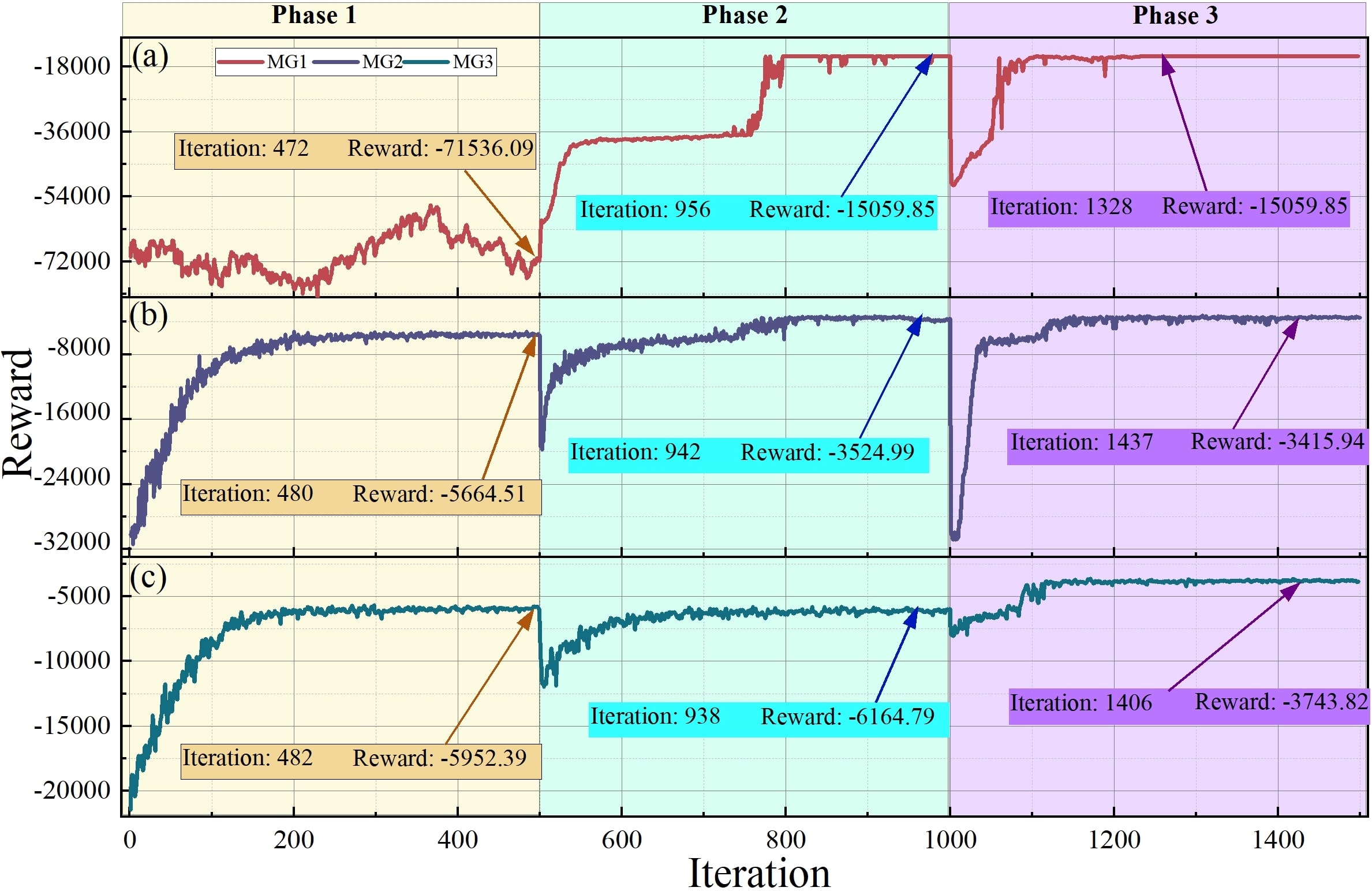}
			\caption{\textcolor{black}{The reward curve of (a) MG1 agent (b) MG2 agent and (c) MG3 agent during the training.}}\label{fig:RewardResult}
		\end{figure}

		In this section, the proposed F-MADRL is applied to the MMG system and its performance evaluation is reported. Fig. \ref{fig:RewardResult} presents the reward curve of the three MG agents, respectively. During the whole training epochs, the FL mechanism is applied every 500 epochs and the training process can be separated into three phases. Since the FL mechanism averages the parameters of the MG agents, the reward value of each agent would dramatically change in the two adjacent phases and all three MG agents receive benefits from this.
		
		For instance, depending on the parameter setting of each MG, the MG1 agent wouldn't converge in the first phase. However, after applying the FL mechanism, its reward raises at the 500th epoch and converges at -15059.85 at the end of phase 2. Besides, although the MG2 agent converges in phase 1, the FL mechanism substantially renews its parameters and further raises the reward to -3524.99 in the second phase. As for the MG3 agent, the benefit of the FL mechanism is mainly shown in the third phase, which helps the agent escape from the local optimal converged at phase 2 and finally achieve a higher reward at the end of the training. In summary, the FL mechanism can be used to get rid of the local optimum, which is caused by insufficient training data due to privacy constraints.
		
		\begin{figure}[htp]
			\centering
			\includegraphics[width=3in]{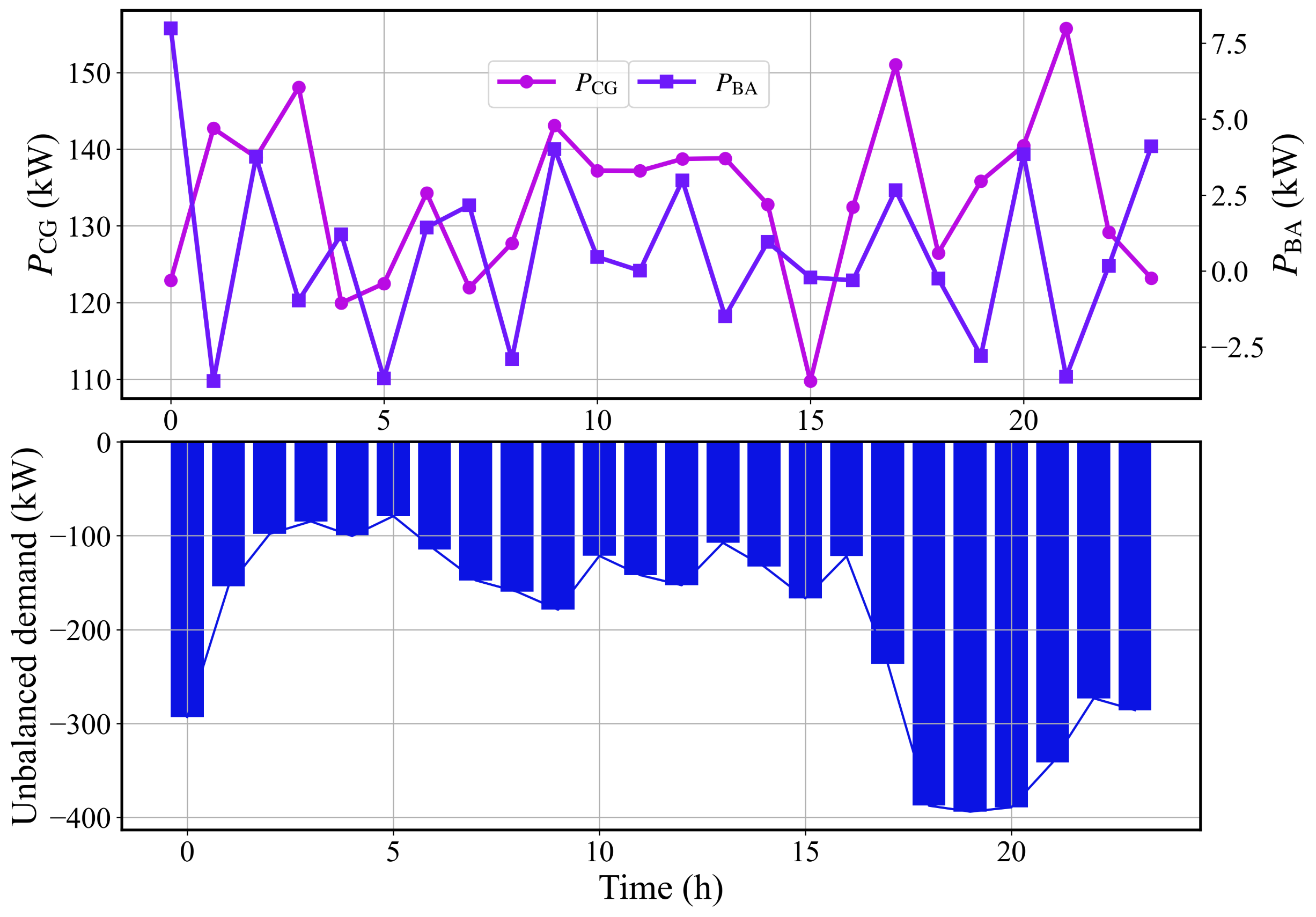}
			\caption{\textcolor{black}{The scheduling of MG1 obtained by the F-MADRL algorithm.}}\label{fig:schedulingMG1}
			\vspace{-0.2cm}
		\end{figure}
		
\begin{figure}[htpb]
			\centering
			\includegraphics[width=3in]{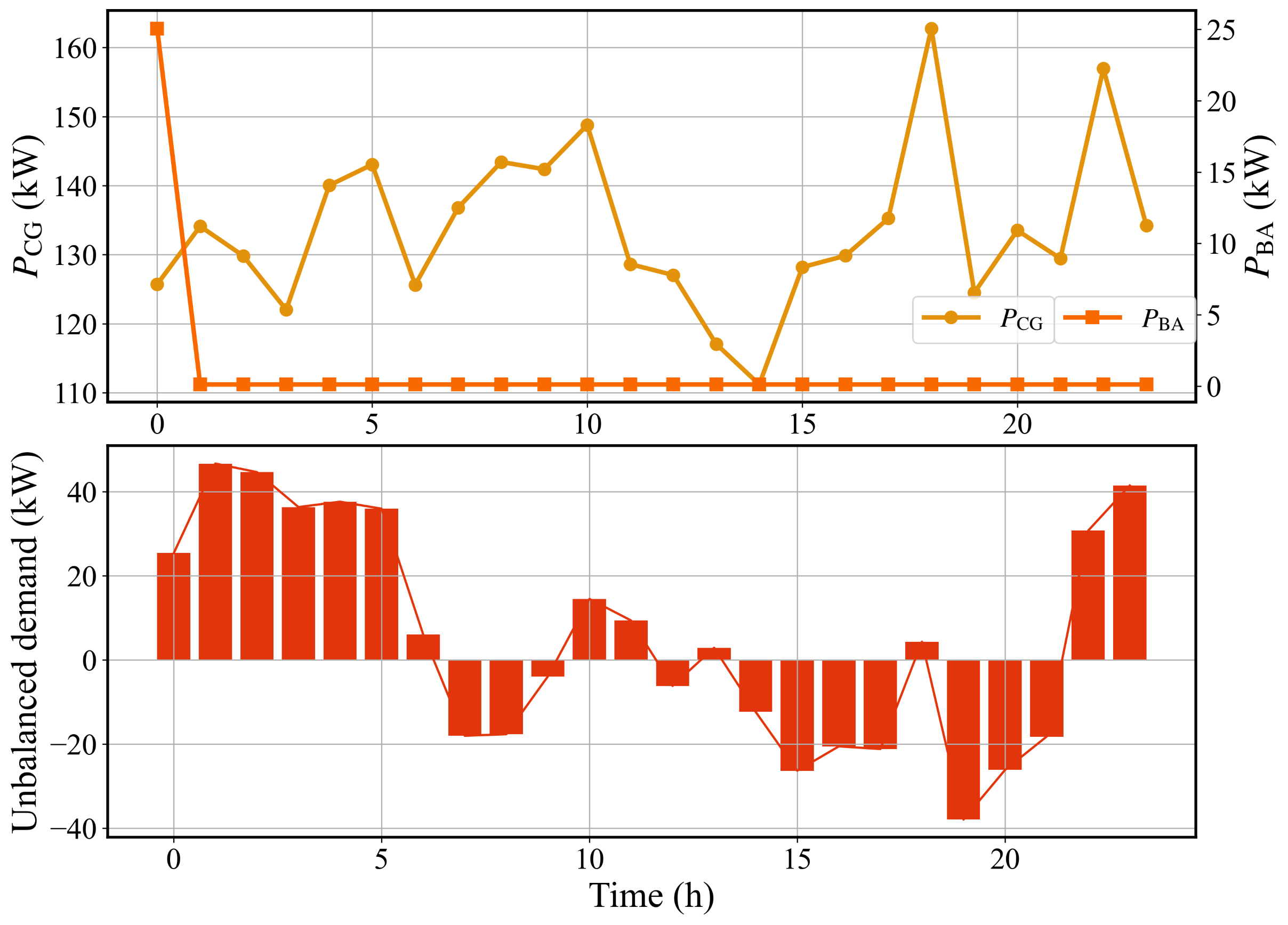}
			\caption{\textcolor{black}{The scheduling of MG3 obtained by the F-MADRL algorithm.}}\label{fig:schedulingMG3}
			\vspace{-0.4cm}
		\end{figure}

		Then, the policies obtained by the F-MADRL are applied to determine the scheduling of each MG. Specifically, Figs. \ref{fig:schedulingMG1}$\sim$\ref{fig:schedulingMG3} denote the scheduling of MG1, MG2 and MG3, respectively. Each figure includes two kinds of graphs, where the above graph is the scheduling solution and the lower graph shows the unbalanced demands, namely the difference between generation and load demands of the MG. The positive unbalanced demands indicate the generation of the MG surpasses its load demands, which means the demands are satisfied while the negative one means the demands are unsatisfied. 
		
		As shown in Fig. \ref{fig:schedulingMG1}, since the demands of MG1 are higher than its capacity, the MG1 operates in the energy self-insufficient state. Therefore, the demands of MG1 are all unsatisfied during the 24 hours. Moreover, the scheduling of BA generation obtained by the agent mainly considers the power balance between its charging and discharging. In the 1st hour, the power demand of MG1 is 457.7 kW, which is higher than the capacity of MG1, thus the agent chooses to discharge the battery for demand supply. Since MG1 operates in the energy self-insufficient state, it requires external power from other MGs and distribution power system, which is achieved by the transaction mechanism in the MMG system. Even the unbalanced demands of MG1 are as high as -402 kW at 20:00, these power shortages can be supplied by other MGs and the distribution power network. This is the reason why the MG1 agent does not fully operate its CG and BA all the time. The agent learns that the energy transaction is more economic than generating power by itself.
		
		\begin{figure}[htpb]
			\centering
			\includegraphics[width=3in]{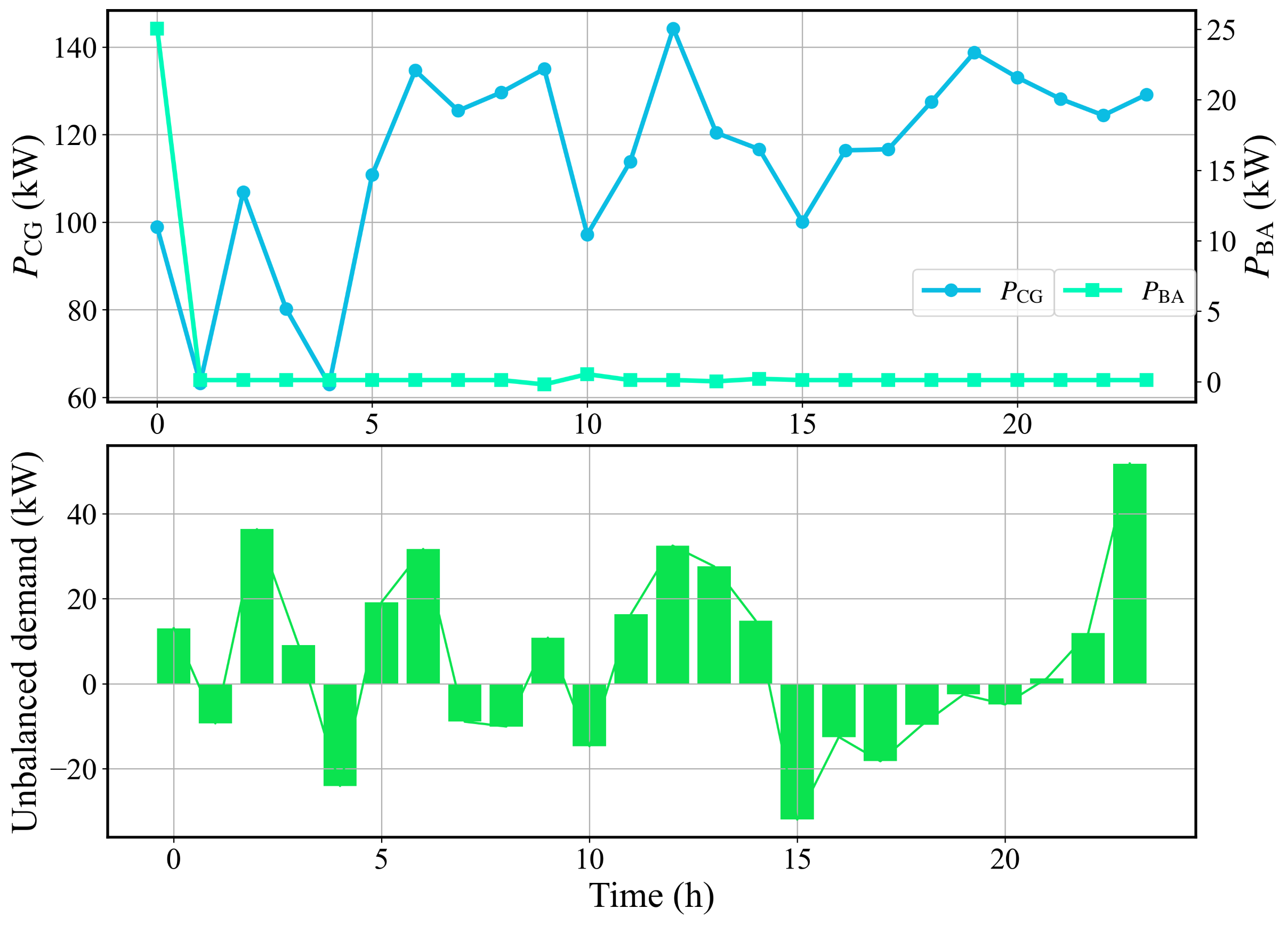}
			\caption{\textcolor{black}{The scheduling of MG2 obtained by the F-MADRL algorithm.}}\label{fig:schedulingMG2}
		\end{figure}
		The scheduling policies of the MG2 agent and MG3 agent are similar but different from that of MG1. \textcolor{black}{As illustrated in Fig. \ref{fig:schedulingMG2} and Fig. \ref{fig:schedulingMG3}, since MG2 and MG3 work in an energy self-sufficient state, their power generations of CG could almost satisfy their demand. In this way, to shrink the generation costs, the two agents perform a similar strategy in the operation of BA. In the most time of the 24 hours, the power outputs of BA are almost 0. Alternatively, the power of CGs would change with the demands of MGs.} For example, in the 15th and 16th hours, the generation and demand of MG2 are nearly equal, thus only causing minor unbalanced demands in these two hours. Besides, the absolute values of unbalanced demands during MG2 and MG3 operation are lower than 50 kW. Those unbalanced demands can be eliminated according to the transaction mechanism in the MMG. In this way, the excess powers are sold to other MGs that work in a self-insufficient state and the power shortages can be supplied by the power from other MGs or the distribution power network.
		
		\textcolor{black}{The above experiments show the efficiency of introducing a federated learning mechanism in the proposed F-MADRL algorithm. The MG agents trained by F-MADRL make efficient scheduling solutions regardless of whether the MG operates in the energy self-sufficient or self-insufficient state.}
		
		{\color{black}\subsection{The Interpretation of the Agents Performance}
			It should be noted that the physics-informed reward would bring about the interpretation of the strategy of agents, to some degree. The three aspects of the reward function, i.e., the costs of CG and BA and the balanced demand, are illustrated along with iterations to study their changes. In this section, six iterations located at a different phase of the training are selected, namely the 1st, 50th, 500th, 700th, 900th and 1400th iterations. Wherein, the 1st and 50th iterations would present the training performance in the early stages, the 500th iteration is located at the end of the first federated phase and the other three iterations are set at the convergent state of the reward value. Note that the values of the reward increase along with the increase of the six iterations.
			\begin{figure}[t]
				\centering
				\includegraphics[width=3.5in]{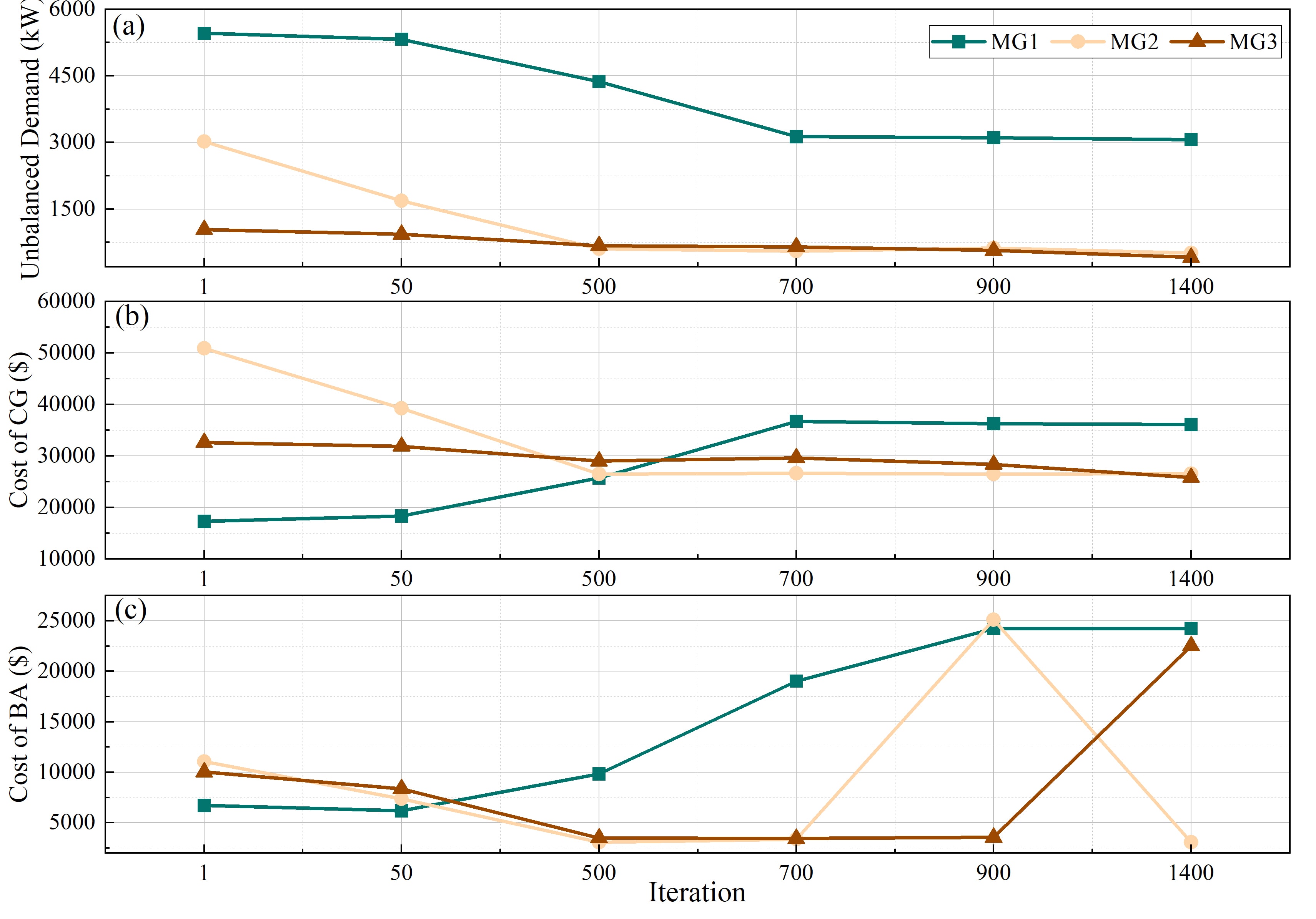}
				\caption{The changes of the decomposition of physics-informed reward according to the training iterations.}\label{fig:rewardDecomposion}
				\vspace{-0.4cm}
			\end{figure}
			
			Fig. \ref{fig:rewardDecomposion} illustrates the changes of the unbalanced demands, costs of CG and cost of BA in the subfigure (a), (b) and (c), respectively. As shown in Fig. \ref{fig:rewardDecomposion}(a), the unbalanced demand of the MG1 decreases from about 5500kW to 3000kW at the 1400th iteration. Besides, the figures for MG2 and MG3 also present a downward trend, they start at 3000kW and 1000kW at 1st iteration and descend to lower than 500kW after 1400 iterations. This means the agents could learn strategies for energy self-sufficiency as much as possible. Besides, Fig. \ref{fig:rewardDecomposion}(b) and (c) present the different strategies of MG agents when operating in the energy self-sufficient and energy self-insufficient state. As illustrated in Fig. \ref{fig:rewardDecomposion}(b), the CG costs of MG2 and MG3 decrease whereas MG1 raises. A similar situation can also be observed in Fig. \ref{fig:rewardDecomposion}(c), the costs of BA for MG1 increase from around \$6000 to \$25000. The BA costs for MG2 and MG3 continuously decrease from about \$10000 and dramatically increase at 900th and 1400th iterations. 
			
			Overall, the MG agents trained by our proposed F-MADRL algorithm could satisfy the target of the MMG system, namely, the energy self-sufficient with the designed physics-informed reward. The algorithm has endowed the explainability by analyzing the performance of the MG agents from the angle of the physics-informed reward.}
		
		\subsection{Performance Comparison}
		\textcolor{black}{In this section, the effectiveness of introducing the FL mechanism is demonstrated by comparing the performance of the proposed F-MADRL with other algorithms that merely implement the self-training of MG agents. These algorithms include numerous well-known deep reinforcement learning algorithms, namely, PPO, A2C and TRPO. Since they are conducted on the multiple MG agents, these comparative algorithms are termed PPO-MADRL, A2C-MADRL and TRPO-MADRL, respectively.} 
		
		The comparisons are conducted from two aspects, i.e., the convergence and the generalization. First, the convergences of the MG agents are illustrated by their reward curves. Besides, to compare the generalization of the MG agents, the testing rewards are set as the metric, which are obtained by testing the agents in both the energy self-sufficient and self-insufficient states. \textcolor{black}{The MG agents are easy to get trapped in local optima since they are trained by the local operation data, which merely contain the perference of the local user. Conesequently, the testing reward is applied to well verify the generalization of the F-MADRL.} \textcolor{black}{Note that the physics-informed reward is a sufficient index for the performance comparison, since the economic cost and power balance are measured in the reward, simultaneously.}
		
		\begin{figure}[htp]
			\centering
			\includegraphics[width=3.5in]{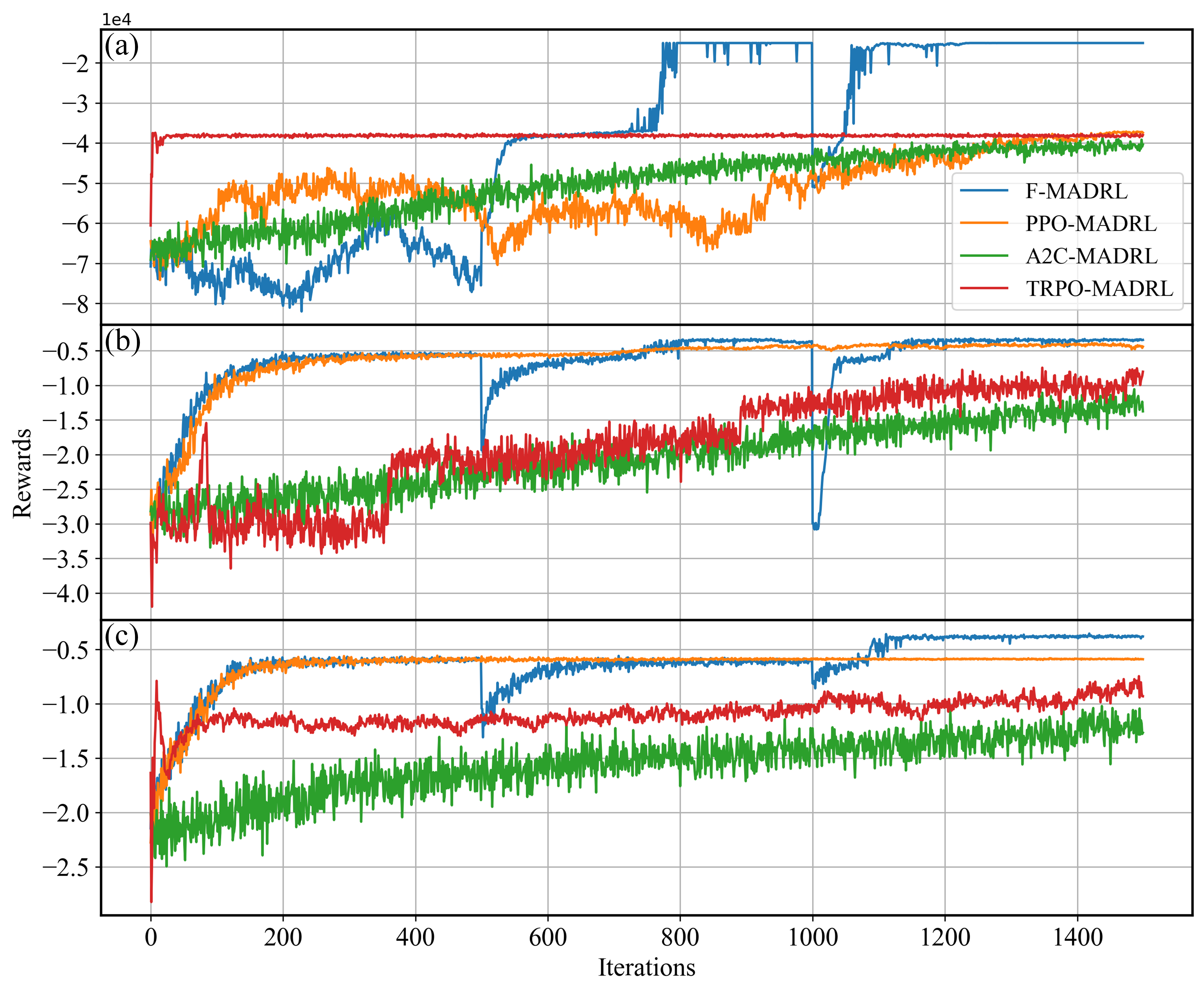}
			\caption{\textcolor{black}{The training reward of each MG agent.}}\label{fig:ComparisonFLDE}
		\end{figure}
		Fig. \ref{fig:ComparisonFLDE} compares the convergence of each algorithm, and the subplots (a), (b) and (c) represent the training process of the MG1 agent, MG2 agent and MG3 agent, respectively. As shown in this figure, the A2C-MADRL performs the worst, under the same learning rate and training epoch, the agents trained by A2C-MADRL cannot be well converged. In addition, although the PPO-MADRL and TRPO-MADRL are able to train the converge agents, their rewards are lower than that of F-MADRL because of lacking the mechanism of sharing information. 
		
		\begin{table}[htbp] 
			\centering
			\caption{The test rewards of the MG agents}
			\scalebox{0.7}{
				\begin{tabular}{cccccc}
					\toprule
					Working State & Agent & F-MADRL & PPO-MADRL & A2C-MADRL & TRPO-MADRL \\
					\midrule
					\multirow{3}[2]{*}{Energy self-sufficient } & MG1   & \textbf{-29017 } & -32727  & -56732  & -46231  \\
					& MG2   & \textbf{-9529 } & -12361  & -19757  & -21818  \\
					& MG3   & \textbf{-2935 } & -5832  & -9782  & -5691  \\
					\midrule
					\multirow{3}[2]{*}{Energy self-insufficient } & MG1   & \textbf{-28650 } & -54881  & -54834  & -29517  \\
					& MG2   & \textbf{-10276 } & -23854  & -22151  & -32756  \\
					& MG3   & \textbf{-3172 } & -5900  & -5005  & -3842  \\
					\bottomrule
			\end{tabular}}%
			\label{tab:generalization_test_reward}%
		\end{table}%
		
		In addition to the reward curves, the generalization of F-MADRL is verified in Table. \ref{tab:generalization_test_reward}, which presents the test rewards of each MG agent in the energy self-sufficient and self-insufficient state. In this figure, the value of the best test reward under each algorithm is bold. \textcolor{black}{It can be learnt that the test rewards obtained by F-MADRL are -29017, -9529 and -2935 for the three MG agents under an energy self-sufficient state, which surpasses other comparative algorithms. Besides, the test rewards of three MG agents trained by F-MADRL are -28650, -10276 and -3172, which perform the best in the energy self-insufficient state along with the comparative algorithms as well.}
		In addition, since the performance of the F-MADRL is better than those of comparative algorithms in both energy self-sufficient and self-insufficient states, it can be concluded that the F-MADRL has a better generalization performance.
		
		\textcolor{black}{The comparisons clarify the introduction of the FL mechanism leads to performance diversity between the proposed F-MADRL and the other three algorithms.} Since the MG agents trained by PPO-MADRL, A2C-MADRL and TRPO-MADRL are only based on the local operation data of MG due to the limitation of privacy and thus causing lower diversity of the training data. Consequently, the decision-making ability of MG agents would decline. The introduction of the FL mechanism alleviates this drawback. By using FL, the experiences of MG agents can be shared without threatening user privacy and data security. In this way, the generalization ability of the agent would be improved, as verified in the above experiments.
		Therefore, the comparisons conducted in this section demonstrate the effectiveness of introducing the FL mechanism in the MADRL algorithm and also reveal a better generalization of F-MADRL.
		
		\section{Conclusion}
		This paper proposes a federated multi-agent deep reinforcement learning algorithm for the multi-microgrids system energy management. \textcolor{black}{A decentralized MMG model is built first, which includes numerous isolated MGs and an agent is used to control the dispatchable elements of each MG to achieve the physics-informed reward.} Due to the privacy protection and data security, the F-MADRL is implemented to train the agents. First, each agent adopts the self-training. \textcolor{black}{Then, the FL mechanism is introduced to build a global agent that aggregates the parameters of all local agents on the server and replaces the local MG agent with the global one.} Therefore, the experiences of each agent can be shared without threatening the privacy and data security.
		
		\textcolor{black}{The case studies are conducted on a MMG with three isolated MGs. The convergence and the performance of F-MADRL are illustrated first. \textcolor{black}{Then, explanations of the strategy of the three MG agents are presented by decomposing the physics-informed reward under different iterations.} Afterwards, by comparing with PPO-MADRL, A2C-MADRL and TRPO-MADRL, the F-MADRL achieves higher test rewards, which means a better generalization. Therefore, it indicates the performance enhancement of introducing the FL mechanism in the MADRL and also demonstrates the effectiveness of our proposed F-MADRL.}
		In this paper, the uncertainty of renewable energy is not considered because its complexity will cause difficulties in the training of F-MADRL and reduce the accuracy of the MG agent strategy. This issue is worth investigating in our future work.

\bibliographystyle{ieeetr}
\bibliography{ref.bib}

\end{document}